\documentclass[12pt,letter,onecolumn]{IEEEtran}
\makeatletter

\let\proof\@undefined
\let\endproof\@undefined
\makeatother

\usepackage[cmex10]{amsmath}
\usepackage{amsfonts}
\usepackage{verbatim}
\usepackage{bm}
\usepackage[pdftex]{color,graphicx}
\usepackage{mdwtab}
\usepackage{subfigure}

\newtheorem{thm}{Theorem}

\newtheorem{prop}[thm]{Proposition}

\newcommand{\defineqq}{\ensuremath{\stackrel{\text{\tiny def}}{=}}}

\begin{document}

\renewcommand{\textfraction}{0}
\title{Hybrid Digital-Analog Codes for\\ Source-Channel Broadcast of \\Gaussian
Sources over Gaussian Channels
}
\author{Vinod M. Prabhakaran,~\IEEEmembership{Member,~IEEE,} Rohit Puri,\\ and
Kannan Ramchandran,~\IEEEmembership{Fellow,~IEEE}
\thanks{Part of this work was presented at Allerton Conference 2005~\cite{prabhakaranprha05}. Vinod M. Prabhakaran is with the Coordinated Science Laboratory, University of Illinois, Urbana-Champaign, Urbana, IL 61801, USA, e-mail:
vinodmp@uiuc.edu. Rohit Puri may be contacted at rpuri@eecs.berkeley.edu.
Kannan Ramchandran is with the Wireless Foundations, Dept. of EECS,
University of California, Berkeley, CA 94702, USA, e-mail:
kannanr@eecs.berkeley.edu.}
}
\maketitle

\begin{abstract}

The problem of broadcasting a parallel Gaussian source over an additive
white Gaussian noise broadcast channel under the mean-squared error
distortion criterion is studied. A hybrid digital-analog coding strategy
which combines source coding with side information, channel coding with
side information, layered source coding, and superposition broadcast
channel coding is presented. When specialized to the open problem of
broadcasting a white Gaussian source over an additive white Gaussian noise
broadcast channel with bandwidth mismatch which has been the subject of
several previous investigations, this coding scheme strictly improves on
the state-of-the-art.

\end{abstract}


\section{Introduction}

In this paper, we study the problem of broadcasting the same source to a
set of receivers over a common channel. The objective is to devise an
encoding strategy (which utilizes a common power and channel bandwidth
resource) to simultaneously deliver different qualities of service
depending on the quality of the channel experienced by the receivers.

For the problem of transmitting a memoryless Gaussian source over a
memoryless additive Gaussian noise point-to-point channel ({\em i.e.}, when
there is only one receiver) operating at the same symbol rate (in other
words, when the memoryless source and channel have the same bandwidth),
Goblick recognized that the uncoded transmission strategy of transmitting
the source samples scaled so as to meet the average encoder power
constraint, followed by the optimal linear minimum mean-squared error
estimation of the source samples from the channel observations at the
receiver, results in the optimal delivered quality (measured in
mean-squared error (MSE) distortion)~\cite{goblicktda65}. Since the
transmitter remains the same irrespective of the channel noise variance,
the same strategy is optimal even when there are multiple receivers ({\em
i.e.}, it is optimal for broadcasting over a memoryless Gaussian broadcast
channel~\cite{coverbc72}).  However, the problem remains open when the
source and channel bandwidths are
mismatched~\cite{shamaivzsl98,mittalphd2002,reznicfzbe06}, or more
generally, when the source has memory~\cite{prabhakaranprha05}.

One obvious digital approach to the bandwidth mismatched problem is to use
the classical separation method of scalable source
coding~\cite{equitzcsr91} followed by a degraded-message-set broadcast
channel coding~\cite{kornermartondmsbc77}.  In this approach, a coarse
source layer is communicated as a common message intended for all users and
a refinement layer is communicated only to some of the users. For the
two-user problem, employing this simple scheme, we make the following
observation (which is proved in the appendix~\ref{app:gap}) 
\begin{prop} \label{prop:intro}
For a memoryless Gaussian source, and a memoryless Gaussian broadcast
channel, the gap in mean-squared error distortion (measured in dB) achieved
by the optimal point-to-point source-channel coder and that achieved by the
separation approach above can be upperbounded by a constant which depends
only on the bandwidth mismatch. In particular,
\begin{align*}
\frac{1}{2}\log_2\left(\frac{D_k}{D_k^\text{optimal}}\right) \leq
\frac{\text{BW}_\text{channel}}{\text{BW}_\text{source}},\qquad k=1,2,
\end{align*}
where $D_1$ and $D_2$ are the MSE distortions incurred by the
receivers 1 and 2 respectively, $D_k^\text{optimal}$ is the optimal
point-to-point distortion to receiver~$k$, $k=1,2$, and
$\text{BW}_\text{source}$ and $\text{BW}_\text{channel}$ are the source and
channel bandwidths respectively.
\end{prop}
The above proposition upperbounds the gap between the trivial lowerbound on
the distortion (namely, the optimal point-to-point distortion) and the
distortion achieved by the separation scheme. The gap to optimality of the
separation scheme for any number of users was studied recently
in~\cite{tiands09}. Better achievable strategies have been proposed by
Mittal and Phamdo in~\cite{mittalphd2002} and for the bandwidth expansion
case (where the bandwidth of the channel is larger than that of the source)
in~\cite{shamaivzsl98} and more recently by Reznic, Feder and Zamir
in~\cite{reznicfzbe06}.  Also, an improved upperbound for the bandwidth
expansion case is available in~\cite{reznicfzbe06}. In this paper, we
consider a slightly more general problem which allows for multiple
independent source components and propose an improved achievable strategy.
When our solution is specialized to the memoryless source and memoryless
channel setting with bandwidth mismatch, it improves the methods proposed
in \cite{shamaivzsl98,mittalphd2002}. We would also like to point out that
for the special case of bandwidth expansion, our scheme essentially matches
the proposal of Reznic, Feder, and Zamir~\cite{reznicfzbe06}. We obtain a
slight improvement over~\cite{reznicfzbe06} through a generalization
overlooked there.

For the case of a memoryless Gaussian source communicated over an additive
memoryless Gaussian noise broadcast channel operating at the same symbol rate,
the observation of Goblick mentioned earlier gives a rather simple optimal
scheme -- transmit the source samples scaled so as to meet the average
encoder power constraint, and the receivers perform the optimal linear
minimum mean-squared error estimation of the source samples from their
respective channel observations. This illustrates an interesting feature of
analog methods -- their ability to enable simultaneous enjoyment of the
power and bandwidth resource by each of the broadcast receivers. 

On the other hand, the obvious digital approach to the above problem is to
use the classical separation method of scalable source coding
\cite{equitzcsr91} followed by degraded message-set broadcast channel
coding \cite{kornermartondmsbc77}.  In this approach, the coarse source
layer is communicated as a common message intended for all users and the
refinement layer is communicated only to some of the users. Consider a
white Gaussian broadcast channel with two receivers; let us call the one
with the lower noise variance {\em strong} and the other {\em weak}. While
the common portion of the information, being limited by the weak receiver,
is sub-optimal for the strong receiver, the refinement portion is
completely unusable by the weak receiver and in fact acts as interference
to it. Thus unlike the analog methods, the digital approach necessarily
involves a ``splitting'' of the total system resource. 

While the above discussion illustrates the power of analog methods,
real-world sources are characterized by a high degree of memory, and thus
they are far from the memoryless model. Parallel source models describe
these sources more effectively than a memoryless model. For this case, in a
point-to-point set-up, analog transmission is sub-optimal in general. For
a parallel Gaussian source with $m$ source components and an additive
memoryless Gaussian noise channel model with an equal number of
sub-channels, the loss in performance of the analog approach with respect
to the digital approach for sufficiently large transmit powers can be shown
to be
\begin{equation}
\frac{\mbox{Analog MSE Distortion}}{\mbox{Digital MSE Distortion}} =
\left(\frac{(\sigma_1+\sigma_2+\ldots+\sigma_m)/m}{(\sigma_1\sigma_2\ldots
\sigma_m)^{1/m}}\right)^{2} \label{eq:AMGM}
\end{equation}
where $\sigma_j^2$ denotes the variance of the $j$-th source
component~\cite{bergertopam67,prabhakaranprupgrade08}. Thus, this gap grows
with the memory of the source and can be arbitrarily large.

This motivates the main question posed in this paper: what is an efficient
way to broadcast parallel Gaussian sources over memoryless two-user
Gaussian broadcast channels? Our solution is driven by aiming to extract
the best of both the analog and the digital worlds. We do this by invoking
a hybrid uncoded-coded strategy, where the coded system uses a combination
of the tools of successive refinement source coding \cite{equitzcsr91},
source coding with side-information or Wyner-Ziv (WZ) coding
\cite{wynerzrf76}, super-position broadcast channel coding
\cite{coverbc72}, and channel coding with side-information or
Gel'fand-Pinsker (GP) coding~\cite{gelfandpcc80} or dirty-paper
coding~\cite{costawd83}.
We would like to point out that this remains an open problem in general and
we present an achievable strategy which constitutes the state-of-the-art to
the best of our knowledge.

In the next section we present the problem setup. We then proceed by first
considering two special cases in section~\ref{hybrid.sec:2sourcechannel}:
(i) when the weak user obtains its point-to-point optimal performance, and
(ii) when the strong user obtains its point-to-point optimal performance.
In the first case, we design a hybrid analog-digital scheme by modifying
the ideas of successive refinement source coding and superposition channel
coding. The idea which was presented in a conference version of this
paper~\cite{prabhakaranprha05} was also independently explored
in~\cite{Lapidoth06} where a point-to-point setting with a memoryless
source and channel without any bandwidth mismatch was considered. In the
second case, we present a hybrid scheme based on the ideas of source coding
with side-information and channel coding with side-information.
Section~\ref{hybrid.sec:general} then considers a scheme which combines all
these ideas to obtain a trade-off between the qualities of reproductions at
the receivers. We take up the special case of memoryless Gaussian source
and memoryless Gaussian broadcast channel with bandwidth mismatch in
section~\ref{hybrid.sec:bandwidthmismatch}. We conclude with some comments
on potential directions of research.

\begin{figure}[htb]
\begin{center}
\scalebox{0.60}{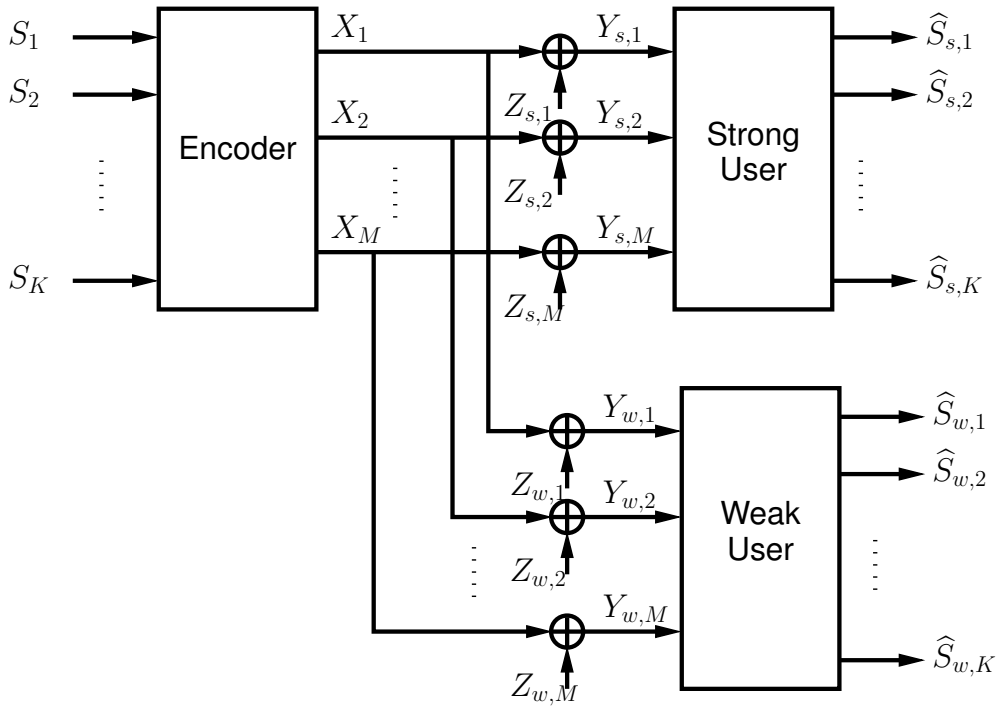}
\label{fig:problemsetup}
\caption{Problem setup. 
}
\end{center}
\end{figure}

\section{Problem setup} \label{hybrid.sec:problemstatement}

The setup is shown in Fig.~\ref{fig:problemsetup}. In this paper, we only
consider the case where all the sub-channels to a particular user have the
same statistics. The upshot of this assumption is a simplification which
results from recognizing that there is an ordering of the users according
to the noise variance of their channel. We will call the user with the
smaller noise variance the {\em strong user}, and the user with the larger
noise variance the {\em weak user}\footnote{It is known that the
performance of such a broadcast channel is identical to that of a {\em
degraded} broadcast channel where the weak user receives the signal
received by the strong user but further corrupted by an additive Gaussian
noise independent of the additive noise corrupting the strong user's
channel and which has a variance equal to the difference between the
variances of the additive noises of the weak user and the strong user in
the original channel. This fact is often expressed by saying that the
original broadcast channel is {\em stochastically degraded}. A proof of the
above stated equivalence for the case of channel coding appears
in~\cite[pg. 422]{covertei91}. The same idea can be used to show an
equivalence for the problem of interest here though we do not need to make
use of this equivalence in our discussion here.}. Note that the ideas
presented in this paper also apply to the case of parallel broadcast
channels where the different sub-channels to the same user may not have the
same statistics, but we do not explore them here.

We let our source have $K$ independent components. The $k$-th component is
denoted by $S_k(i)$ and it is independent and identically distributed
(i.i.d.) over the time index $i=1,2,\ldots$. Our source is Gaussian,
$S_k(i) \sim {\mathcal N}(0,\sigma_k^2)$.  Without loss of generality, we
will assume that $\sigma_1^2 \geq \sigma_2^2 \geq \ldots \geq \sigma_K^2$. 
We will explicitly model the fact that the source and channel bandwidths do
not necessarily match.  We will assume that there are $M$ parallel
broadcast sub-channels of the same statistics.  When $M=K$, the bandwidth
of the source matches the bandwidth of the channel. The weak user observes
${Y_w}_m(i)=X_m(i) + {Z_w}_m(i),\; m=1,2,\ldots,M,\; i = 1,2,\ldots$, and
the strong user observes ${Y_s}_m(i)=X_m(i) + {Z_s}_m(i)$, where $X_m(i)$
is the input to the $m$-th sub-channel. The noise processes are
${Z_w}_m(i)$ and ${Z_s}_m(i)$ independent i.i.d. (over $m$ and $i$)
Gaussians with variances $N_w$ and $N_s$ respectively, where $N_s < N_w$.
The source-channel encoder $f^n:{\mathbb R}^{Kn} \rightarrow {\mathbb
R}^{Mn}$ maps an $n$-length block of the source to an $n$-length block of
the channel input.  There is an average power constraint on the encoder so
that $\left(\sum_{i=1}^n\sum_{m=1}^M X^2_m(i)\right)/(nM) \le P$. The
source-channel decoders $g^n_s:{\mathbb R}^{Mn} \rightarrow {\mathbb
R}^{Kn}$ and $g^n_w:{\mathbb R}^{Mn} \rightarrow {\mathbb R}^{Kn}$ at the
strong and weak user, respectively, reconstruct $n$-length blocks,
\begin{align*}
\left\{\widehat{S_s}_k(i),\;i=1,\ldots,n,k=1,\ldots,K\right\}
&=
g^n_s\left(\left\{{Y_s}_m(i),\;i=1,2,\ldots,n,\;m=1,2,\ldots,M\right\}\right), 
\mbox{and}\\
\left\{\widehat{S_w}_k(i),\;i=1,\ldots,n,k=1,\ldots,K\right\}
&=
g^n_w\left(\left\{{Y_w}_m(i),\;i=1,2,\ldots,n,\;m=1,2,\ldots,M\right\}\right)
\end{align*}
of the source from $n$-length blocks of the channel outputs. Distortions are
measured as the average of the mean-squared error distortion over all
source components
\[D^n_j=\frac{1}{nK}\sum_{i=1}^n
\sum_{k=1}^K\left(S_k(i)-\widehat{S_j}_k(i)\right)^2, \;j\in\{s,w\}.\]
A pair of distortions $(D_s,D_w)$ will be said to be {\em achievable} if for
any $\epsilon>0$, for sufficiently large $n$, there is $(f^n,g_s^n,g_w^n)$ such
that $D_j^n\leq D_j + \epsilon, \;j\in\{s,w\}$. The problem is to characterize
the region of all the achievable distortions $(D_s,D_w)$ for a given transmit
power $P$. This remains open. In the next sections, we present our inner bound
to the region ({\em i.e.}, an achievable region).

\section{An achievable solution: the extreme points} \label{hybrid.sec:2sourcechannel}

In this section we present all the key ideas involved in our achievable
strategy. We consider two extreme cases -- when the weak user achieves its
point-to-point optimal quality, and when the strong user achieves its
point-to-point optimal quality -- to illustrate two complementary
strategies which together constitute the general achievable solution.

\subsection{Weak-user-optimal case}

From Shannon's separation theorem~\cite[pg.~216]{covertei91} we know that
the optimal solution for the point-to-point source-channel problem can be
obtained by the separation principle of first optimally source coding and
then transmitting the resulting bit stream using an optimal channel code.
Thus the lowest distortion $D_w^\ast$ attainable by the weak user is given
by ``reverse water-filling'' over the source
components~\cite[pg.~348]{covertei91}

\begin{equation}
D_w^\ast = \frac{1}{K} \sum_{k=1}^K D_k, \mbox{   where }
   D_k=\left\{ \begin{array}{ll} 
                 \mu, &\mbox{if } \mu < \sigma_k^2,\\
                 \sigma_k^2, &\mbox{if } \mu \ge \sigma_k^2,
               \end{array} \right. \label{hybrid.eq:reversewaterfill}
\end{equation}
where $\mu$ is chosen such that the total rate $(1/2)\sum_{k=1}^K
\log(\sigma_k^2/D_k)$ equals the capacity $C_w$ of the weak user's channel.
$C_w$ is in turn given by $(1/2)\sum_{m=1}^M\log(1+P_m/{N_w})$ where we choose
$P_1=P_2=\ldots=P_M=P$.

\begin{figure}[htb]
\begin{center}
\scalebox{0.8}{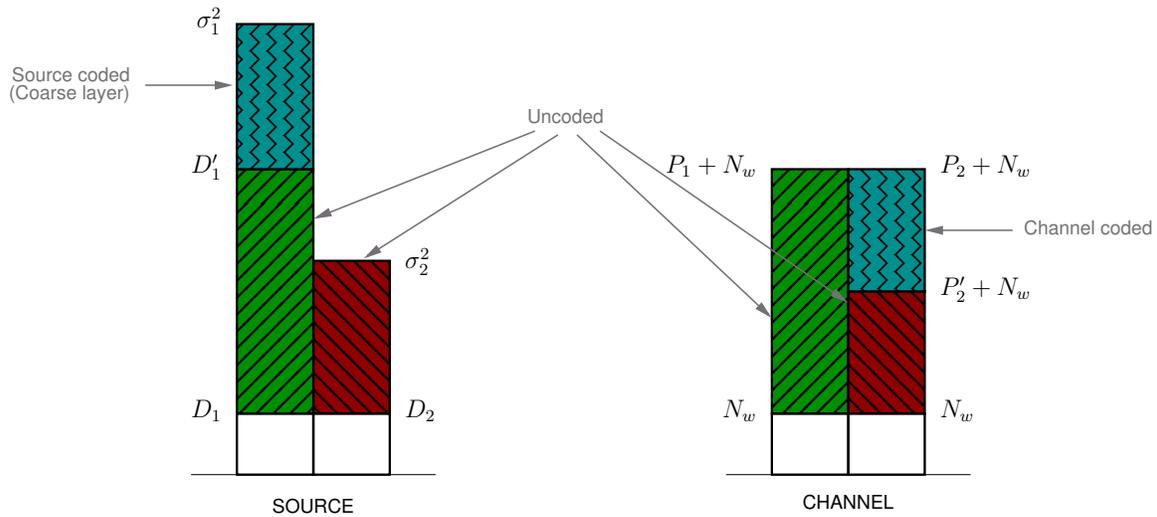}\\
\end{center}
\caption{
Weak-user-optimal case: The schematic diagram shows the allocation for a
$K=M=2$ example. The optimal coded separation scheme for the weak-user may
be thought of as sending source-coded bits about the second source
component $S_2$ and the coarse-layer bits from a successive refinement
source coding of the first source component $S_1$ over the second
sub-channel using superposition coding, and the refinement-layer bits from
the successive refinement coding of $S_1$ alone over
the first sub-channel. An equivalent distortion performance can be achieved
at the weak user while improving the strong user's performance by (i)
sending the quantization error from the coarse quantization of $S_1$
uncoded (scaled) over the first sub-channel, and (ii) sending $S_2$ uncoded
(scaled) over the second sub-channel and the coarse-layer bits on $S_1$
channel coded with sufficient power to be decoded (by the weak user)
treating the uncoded transmission of $S_2$ as noise. The strong user
benefits from both the uncoded transmissions since it can form better
quality estimates than the weak user.}
\label{fig:wuhistogram}
\end{figure}

If this separation strategy is followed for the broadcast case as well, the
strong user also recovers the source at a distortion $D_w^\ast$. 
However, without compromising the quality of reproduction for the weak
user, better quality can be delivered to the strong user. Before presenting
our solution in full generality, it is useful to consider the special case
of $K=M=2$ and $\sigma_1^2 > \sigma_2^2$; see Fig.~\ref{fig:wuhistogram}.
Let us suppose that the optimal point-to-point reverse water-filling
solution for the weak user allocates distortions $D_1$ and $D_2$ for the
source components $S_1$ and $S_2$, respectively. Also,
let us denote the powers allocated to the sub-channels $X_1$ and $X_2$ by
$P_1$ and $P_2$ respectively. $P_1=P_2=P$. Then
\[
\frac{1}{2}\log\left(\frac{\sigma_1^2}{D_1}\frac{\sigma_2^2}{D_2}\right) =
\frac{1}{2}\log\left(\frac{P_1+{N_w}}{{N_w}}\frac{P_2+{N_w}}{{N_w}}\right).\]
We can source code $S_1$ using a successive refinement strategy thereby
producing two bit streams: a coarse description at distortion $D_1^\prime$ and
a refinement stream which refines from $D_1^\prime$ to $D_1$. Since Gaussian
sources are successively refinable~\cite{equitzcsr91}, this can be done
without loss of optimality for the weak user. We choose $D_1^\prime$ such that
the bitrate of the refinement stream is equal to the rate at which the first
sub-channel operates. {\em i.e.},
\[ \frac{1}{2}\log\left(\frac{D_1^\prime}{D_1}\right) =
\frac{1}{2}\log\left(1+\frac{P_1}{{N_w}}\right).\]
Combining the two equations above gives
\[ 
\frac{1}{2}\log\left(\frac{\sigma_1^2}{D_1^\prime}\frac{\sigma_2^2}{D_2}\right)
   = \frac{1}{2}\log\left(1+\frac{P_2}{{N_w}}\right).\]

\begin{figure}[htb]
\begin{center}
\scalebox{0.95}{\setlength{\unitlength}{1184sp}%
\begingroup\makeatletter\ifx\SetFigFont\undefined%
\gdef\SetFigFont#1#2#3#4#5{%
  \reset@font\fontsize{#1}{#2pt}%
  \fontfamily{#3}\fontseries{#4}\fontshape{#5}%
  \selectfont}%
\fi\endgroup%
\begin{picture}(23782,8968)(1166,-20964)
{\color[rgb]{0,0,0}\thicklines
\put(1986,-14647){\circle{472}}
}%
{\color[rgb]{0,0,0}\put(1986,-14497){\line( 0,-1){301}}
}%
{\color[rgb]{0,0,0}\put(1837,-14647){\line( 1, 0){298}}
}%
{\color[rgb]{0,0,0}\put(6065,-13907){\framebox(1842,1889){}}
}%
\put(6962,-12868){\makebox(0,0)[b]{\smash{{\SetFigFont{7}{8.4}{\sfdefault}{\mddefault}{\updefault}{\color[rgb]{0,0,0}channel}%
}}}}
\put(6962,-13388){\makebox(0,0)[b]{\smash{{\SetFigFont{7}{8.4}{\sfdefault}{\mddefault}{\updefault}{\color[rgb]{0,0,0}encoder}%
}}}}
{\color[rgb]{0,0,0}\put(12254,-12969){\circle{472}}
}%
{\color[rgb]{0,0,0}\put(12254,-12819){\line( 0,-1){300}}
}%
{\color[rgb]{0,0,0}\put(12104,-12969){\line( 1, 0){299}}
}%
{\color[rgb]{0,0,0}\put(6065,-20049){\framebox(1842,1890){}}
}%
\put(6962,-19010){\makebox(0,0)[b]{\smash{{\SetFigFont{7}{8.4}{\sfdefault}{\mddefault}{\updefault}{\color[rgb]{0,0,0}channel}%
}}}}
\put(6962,-19529){\makebox(0,0)[b]{\smash{{\SetFigFont{7}{8.4}{\sfdefault}{\mddefault}{\updefault}{\color[rgb]{0,0,0}encoder}%
}}}}
{\color[rgb]{0,0,0}\put(6065,-17451){\framebox(1842,1890){}}
}%
\put(6962,-16411){\makebox(0,0)[b]{\smash{{\SetFigFont{7}{8.4}{\sfdefault}{\mddefault}{\updefault}{\color[rgb]{0,0,0}channel}%
}}}}
\put(6962,-16931){\makebox(0,0)[b]{\smash{{\SetFigFont{7}{8.4}{\sfdefault}{\mddefault}{\updefault}{\color[rgb]{0,0,0}encoder}%
}}}}
{\color[rgb]{0,0,0}\put(2994,-20049){\framebox(1842,1890){}}
}%
\put(3891,-19010){\makebox(0,0)[b]{\smash{{\SetFigFont{7}{8.4}{\sfdefault}{\mddefault}{\updefault}{\color[rgb]{0,0,0}source}%
}}}}
\put(3891,-19529){\makebox(0,0)[b]{\smash{{\SetFigFont{7}{8.4}{\sfdefault}{\mddefault}{\updefault}{\color[rgb]{0,0,0}encoder}%
}}}}
{\color[rgb]{0,0,0}\put(15057,-17704){\circle{472}}
}%
{\color[rgb]{0,0,0}\put(15057,-17553){\line( 0,-1){301}}
}%
{\color[rgb]{0,0,0}\put(14908,-17704){\line( 1, 0){298}}
}%
{\color[rgb]{0,0,0}\put(20002,-20049){\framebox(1842,1890){}}
}%
\put(20899,-19010){\makebox(0,0)[b]{\smash{{\SetFigFont{7}{8.4}{\sfdefault}{\mddefault}{\updefault}{\color[rgb]{0,0,0}source}%
}}}}
\put(20899,-19529){\makebox(0,0)[b]{\smash{{\SetFigFont{7}{8.4}{\sfdefault}{\mddefault}{\updefault}{\color[rgb]{0,0,0}decoder}%
}}}}
{\color[rgb]{0,0,0}\put(16269,-20049){\framebox(1843,1890){}}
}%
\put(17167,-19010){\makebox(0,0)[b]{\smash{{\SetFigFont{7}{8.4}{\sfdefault}{\mddefault}{\updefault}{\color[rgb]{0,0,0}channel}%
}}}}
\put(17167,-19529){\makebox(0,0)[b]{\smash{{\SetFigFont{7}{8.4}{\sfdefault}{\mddefault}{\updefault}{\color[rgb]{0,0,0}decoder}%
}}}}
{\color[rgb]{0,0,0}\put(12211,-16569){\circle{472}}
}%
{\color[rgb]{0,0,0}\put(12211,-16419){\line( 0,-1){301}}
}%
{\color[rgb]{0,0,0}\put(12062,-16569){\line( 1, 0){298}}
}%
{\color[rgb]{0,0,0}\put(9550,-16547){\circle{472}}
}%
{\color[rgb]{0,0,0}\put(9550,-16397){\line( 0,-1){301}}
}%
{\color[rgb]{0,0,0}\put(9401,-16547){\line( 1, 0){298}}
}%
{\color[rgb]{0,0,0}\put(9782,-16563){\vector( 1, 0){2193}}
}%
{\color[rgb]{0,0,0}\put(12436,-16569){\vector( 1, 0){3838}}
}%
{\color[rgb]{0,0,0}\put(7939,-16563){\vector( 1, 0){1343}}
}%
\put(12353,-18500){\makebox(0,0)[b]{\smash{{\SetFigFont{7}{8.4}{\familydefault}{\mddefault}{\updefault}{\color[rgb]{0,0,0}$Z_{w_2}$}%
}}}}
\put(3913,-13105){\makebox(0,0)[b]{\smash{{\SetFigFont{7}{8.4}{\sfdefault}{\mddefault}{\updefault}{\color[rgb]{0,0,0}source}%
}}}}
\put(3913,-13625){\makebox(0,0)[b]{\smash{{\SetFigFont{7}{8.4}{\sfdefault}{\mddefault}{\updefault}{\color[rgb]{0,0,0}encoder}%
}}}}
\put(3913,-12680){\makebox(0,0)[b]{\smash{{\SetFigFont{7}{8.4}{\sfdefault}{\mddefault}{\updefault}{\color[rgb]{0,0,0}SR}%
}}}}
\put(3913,-16280){\makebox(0,0)[b]{\smash{{\SetFigFont{7}{8.4}{\sfdefault}{\mddefault}{\updefault}{\color[rgb]{0,0,0}source}%
}}}}
\put(3913,-16800){\makebox(0,0)[b]{\smash{{\SetFigFont{7}{8.4}{\sfdefault}{\mddefault}{\updefault}{\color[rgb]{0,0,0}encoder}%
}}}}
{\color[rgb]{0,0,0}\put(16269,-13907){\framebox(1843,1889){}}
}%
\put(17167,-12868){\makebox(0,0)[b]{\smash{{\SetFigFont{7}{8.4}{\sfdefault}{\mddefault}{\updefault}{\color[rgb]{0,0,0}channel}%
}}}}
\put(17167,-13388){\makebox(0,0)[b]{\smash{{\SetFigFont{7}{8.4}{\sfdefault}{\mddefault}{\updefault}{\color[rgb]{0,0,0}decoder}%
}}}}
{\color[rgb]{0,0,0}\put(23246,-16522){\circle{472}}
}%
{\color[rgb]{0,0,0}\put(23246,-16372){\line( 0,-1){301}}
}%
{\color[rgb]{0,0,0}\put(23097,-16522){\line( 1, 0){298}}
}%
\put(20931,-12675){\makebox(0,0)[b]{\smash{{\SetFigFont{7}{8.4}{\sfdefault}{\mddefault}{\updefault}{\color[rgb]{0,0,0}SR}%
}}}}
\put(20931,-13100){\makebox(0,0)[b]{\smash{{\SetFigFont{7}{8.4}{\sfdefault}{\mddefault}{\updefault}{\color[rgb]{0,0,0}source}%
}}}}
\put(20931,-13620){\makebox(0,0)[b]{\smash{{\SetFigFont{7}{8.4}{\sfdefault}{\mddefault}{\updefault}{\color[rgb]{0,0,0}decoder}%
}}}}
\put(20908,-16383){\makebox(0,0)[b]{\smash{{\SetFigFont{7}{8.4}{\sfdefault}{\mddefault}{\updefault}{\color[rgb]{0,0,0}source}%
}}}}
\put(20908,-16903){\makebox(0,0)[b]{\smash{{\SetFigFont{7}{8.4}{\sfdefault}{\mddefault}{\updefault}{\color[rgb]{0,0,0}decoder}%
}}}}
{\color[rgb]{0,0,0}\put(5122,-14616){\vector(-1, 0){2884}}
\put(5122,-14615){\line( 0,-1){1905}}
}%
{\color[rgb]{0,0,0}\put(4864,-12962){\vector( 1, 0){1201}}
}%
{\color[rgb]{0,0,0}\put(7907,-12962){\vector( 1, 0){4110}}
}%
{\color[rgb]{0,0,0}\put(12478,-12969){\vector( 1, 0){3791}}
}%
{\color[rgb]{0,0,0}\put(12254,-14569){\vector( 0, 1){1375}}
}%
{\color[rgb]{0,0,0}}%
{\color[rgb]{0,0,0}\put(4864,-16506){\vector( 1, 0){1201}}
}%
{\color[rgb]{0,0,0}\put(4864,-19104){\vector( 1, 0){1201}}
}%
{\color[rgb]{0,0,0}\put(2994,-17451){\framebox(1842,1890){}}
}%
{\color[rgb]{0,0,0}\put(1293,-19104){\vector( 1, 0){1701}}
}%
{\color[rgb]{0,0,0}\put(18139,-19104){\vector( 1, 0){1863}}
}%
{\color[rgb]{0,0,0}\put(18584,-17687){\vector(-1, 0){3307}}
\put(18584,-17687){\line( 0, 1){1197}}
}%
{\color[rgb]{0,0,0}}%
{\color[rgb]{0,0,0}\put(21871,-19104){\vector( 1, 0){3044}}
}%
{\color[rgb]{0,0,0}\put(21844,-12962){\line( 1, 0){1417}}
\put(23261,-12962){\vector( 0,-1){3307}}
}%
{\color[rgb]{0,0,0}\put(18092,-16506){\vector( 1, 0){1910}}
}%
{\color[rgb]{0,0,0}\put(21871,-16506){\vector( 1, 0){1154}}
}%
{\color[rgb]{0,0,0}\put(23478,-16506){\vector( 1, 0){1437}}
}%
{\color[rgb]{0,0,0}\put(16269,-17451){\framebox(1843,1890){}}
}%
{\color[rgb]{0,0,0}\put(12211,-18169){\vector( 0, 1){1374}}
}%
{\color[rgb]{0,0,0}\put(7934,-19104){\line( 1, 0){1605}}
\put(9539,-19100){\vector( 0, 1){2310}}
}%
{\color[rgb]{0,0,0}\put(15037,-16580){\vector( 0,-1){892}}
}%
{\color[rgb]{0,0,0}\put(15051,-19115){\vector( 1, 0){1223}}
\put(15051,-19115){\line( 0, 1){1145}}
}%
{\color[rgb]{0,0,0}\put(2994,-13907){\framebox(1842,1889){}}
}%
{\color[rgb]{0,0,0}\put(2002,-16580){\vector( 0, 1){1681}}
}%
{\color[rgb]{0,0,0}\put(2002,-12995){\vector( 1, 0){989}}
\put(2002,-12988){\line( 0,-1){1439}}
}%
{\color[rgb]{0,0,0}\put(1199,-16573){\vector( 1, 0){1748}}
}%
{\color[rgb]{0,0,0}\put(20002,-13907){\framebox(1842,1889){}}
}%
{\color[rgb]{0,0,0}\put(18092,-12962){\vector( 1, 0){1910}}
}%
{\color[rgb]{0,0,0}\put(20002,-17451){\framebox(1842,1890){}}
}%
\put(1830,-18909){\makebox(0,0)[b]{\smash{{\SetFigFont{7}{8.4}{\familydefault}{\mddefault}{\updefault}{\color[rgb]{0,0,0}$S_2$}%
}}}}
\put(2521,-14474){\makebox(0,0)[b]{\smash{{\SetFigFont{7}{8.4}{\familydefault}{\mddefault}{\updefault}{\color[rgb]{0,0,0}$-$}%
}}}}
\put(24206,-16364){\makebox(0,0)[b]{\smash{{\SetFigFont{7}{8.4}{\sfdefault}{\mddefault}{\updefault}{\color[rgb]{0,0,0}$\widehat{S}_1$}%
}}}}
\put(24112,-18962){\makebox(0,0)[b]{\smash{{\SetFigFont{7}{8.4}{\sfdefault}{\mddefault}{\updefault}{\color[rgb]{0,0,0}$\widehat{S}_2$}%
}}}}
\put(8616,-12726){\makebox(0,0)[b]{\smash{{\SetFigFont{7}{8.4}{\familydefault}{\mddefault}{\updefault}{\color[rgb]{0,0,0}$(P_1)$}%
}}}}
\put(15561,-17592){\makebox(0,0)[b]{\smash{{\SetFigFont{7}{8.4}{\familydefault}{\mddefault}{\updefault}{\color[rgb]{0,0,0}$-$}%
}}}}
\put(16647,-20805){\makebox(0,0)[b]{\smash{{\SetFigFont{7}{8.4}{\sfdefault}{\mddefault}{\updefault}{\color[rgb]{0.333,0.333,0.333}Superposition Decoder}%
}}}}
\put(7624,-20805){\makebox(0,0)[b]{\smash{{\SetFigFont{7}{8.4}{\sfdefault}{\mddefault}{\updefault}{\color[rgb]{0.333,0.333,0.333}Superposition Encoder}%
}}}}
\put(17167,-16411){\makebox(0,0)[b]{\smash{{\SetFigFont{7}{8.4}{\sfdefault}{\mddefault}{\updefault}{\color[rgb]{0,0,0}channel}%
}}}}
\put(17167,-16931){\makebox(0,0)[b]{\smash{{\SetFigFont{7}{8.4}{\sfdefault}{\mddefault}{\updefault}{\color[rgb]{0,0,0}decoder}%
}}}}
\put(12395,-14899){\makebox(0,0)[b]{\smash{{\SetFigFont{7}{8.4}{\familydefault}{\mddefault}{\updefault}{\color[rgb]{0,0,0}$Z_{w_1}$}%
}}}}
\put(13032,-16325){\makebox(0,0)[b]{\smash{{\SetFigFont{7}{8.4}{\familydefault}{\mddefault}{\updefault}{\color[rgb]{0,0,0}$Y_{w_2}$}%
}}}}
\put(11143,-16325){\makebox(0,0)[b]{\smash{{\SetFigFont{7}{8.4}{\familydefault}{\mddefault}{\updefault}{\color[rgb]{0,0,0}$X_2$}%
}}}}
\put(11175,-12733){\makebox(0,0)[b]{\smash{{\SetFigFont{7}{8.4}{\familydefault}{\mddefault}{\updefault}{\color[rgb]{0,0,0}$X_1$}%
}}}}
\put(8861,-19718){\makebox(0,0)[b]{\smash{{\SetFigFont{7}{8.4}{\familydefault}{\mddefault}{\updefault}{\color[rgb]{0,0,0}$(P_2^{\prime})$}%
}}}}
\put(8826,-16197){\makebox(0,0)[b]{\smash{{\SetFigFont{7}{8.4}{\familydefault}{\mddefault}{\updefault}{\color[rgb]{0,0,0}$(P_2-P_2^{\prime})$}%
}}}}
\put(13075,-12724){\makebox(0,0)[b]{\smash{{\SetFigFont{7}{8.4}{\familydefault}{\mddefault}{\updefault}{\color[rgb]{0,0,0}$Y_{w_1}$}%
}}}}
\put(1783,-17098){\makebox(0,0)[b]{\smash{{\SetFigFont{7}{8.4}{\familydefault}{\mddefault}{\updefault}{\color[rgb]{0,0,0}$S_1$}%
}}}}
\end{picture}%
}\\
(a)\\
\scalebox{0.95}{\setlength{\unitlength}{1184sp}%
\begingroup\makeatletter\ifx\SetFigFont\undefined%
\gdef\SetFigFont#1#2#3#4#5{%
  \reset@font\fontsize{#1}{#2pt}%
  \fontfamily{#3}\fontseries{#4}\fontshape{#5}%
  \selectfont}%
\fi\endgroup%
\begin{picture}(23782,8068)(1166,-20074)
{\color[rgb]{0,0,0}\thicklines
\put(1986,-14647){\circle{472}}
}%
{\color[rgb]{0,0,0}\put(1986,-14497){\line( 0,-1){301}}
}%
{\color[rgb]{0,0,0}\put(1837,-14647){\line( 1, 0){298}}
}%
{\color[rgb]{0,0,0}\put(12254,-12969){\circle{472}}
}%
{\color[rgb]{0,0,0}\put(12254,-12819){\line( 0,-1){300}}
}%
{\color[rgb]{0,0,0}\put(12104,-12969){\line( 1, 0){299}}
}%
{\color[rgb]{0,0,0}\put(6065,-17451){\framebox(1842,1890){}}
}%
\put(6962,-16411){\makebox(0,0)[b]{\smash{{\SetFigFont{7}{8.4}{\sfdefault}{\mddefault}{\updefault}{\color[rgb]{0,0,0}channel}%
}}}}
\put(6962,-16931){\makebox(0,0)[b]{\smash{{\SetFigFont{7}{8.4}{\sfdefault}{\mddefault}{\updefault}{\color[rgb]{0,0,0}encoder}%
}}}}
{\color[rgb]{0,0,0}\put(15057,-17704){\circle{472}}
}%
{\color[rgb]{0,0,0}\put(15057,-17553){\line( 0,-1){301}}
}%
{\color[rgb]{0,0,0}\put(14908,-17704){\line( 1, 0){298}}
}%
{\color[rgb]{0,0,0}\put(12211,-16569){\circle{472}}
}%
{\color[rgb]{0,0,0}\put(12211,-16419){\line( 0,-1){301}}
}%
{\color[rgb]{0,0,0}\put(12062,-16569){\line( 1, 0){298}}
}%
{\color[rgb]{0,0,0}\put(9550,-16547){\circle{472}}
}%
{\color[rgb]{0,0,0}\put(9550,-16397){\line( 0,-1){301}}
}%
{\color[rgb]{0,0,0}\put(9401,-16547){\line( 1, 0){298}}
}%
{\color[rgb]{0,0,0}\put(9782,-16563){\vector( 1, 0){2193}}
}%
{\color[rgb]{0,0,0}\put(12436,-16569){\vector( 1, 0){3838}}
}%
{\color[rgb]{0,0,0}\put(7939,-16563){\vector( 1, 0){1343}}
}%
\put(12353,-18500){\makebox(0,0)[b]{\smash{{\SetFigFont{7}{8.4}{\familydefault}{\mddefault}{\updefault}{\color[rgb]{0,0,0}$Z_{w_2}$}%
}}}}
\put(3913,-16280){\makebox(0,0)[b]{\smash{{\SetFigFont{7}{8.4}{\sfdefault}{\mddefault}{\updefault}{\color[rgb]{0,0,0}source}%
}}}}
\put(3913,-16800){\makebox(0,0)[b]{\smash{{\SetFigFont{7}{8.4}{\sfdefault}{\mddefault}{\updefault}{\color[rgb]{0,0,0}encoder}%
}}}}
{\color[rgb]{0,0,0}\put(23246,-16522){\circle{472}}
}%
{\color[rgb]{0,0,0}\put(23246,-16372){\line( 0,-1){301}}
}%
{\color[rgb]{0,0,0}\put(23097,-16522){\line( 1, 0){298}}
}%
\put(20908,-16383){\makebox(0,0)[b]{\smash{{\SetFigFont{7}{8.4}{\sfdefault}{\mddefault}{\updefault}{\color[rgb]{0,0,0}source}%
}}}}
\put(20908,-16903){\makebox(0,0)[b]{\smash{{\SetFigFont{7}{8.4}{\sfdefault}{\mddefault}{\updefault}{\color[rgb]{0,0,0}decoder}%
}}}}
{\color[rgb]{0,0,0}\put(4127,-20052){\framebox(2551,1889){}}
}%
\put(5402,-19013){\makebox(0,0)[b]{\smash{{\SetFigFont{7}{8.4}{\sfdefault}{\mddefault}{\updefault}{\color[rgb]{0,0,0}Power}%
}}}}
\put(5402,-19533){\makebox(0,0)[b]{\smash{{\SetFigFont{7}{8.4}{\sfdefault}{\mddefault}{\updefault}{\color[rgb]{0,0,0}scaling}%
}}}}
{\color[rgb]{0,0,0}\put(17402,-20052){\framebox(2551,1889){}}
}%
\put(18678,-19013){\makebox(0,0)[b]{\smash{{\SetFigFont{7}{8.4}{\sfdefault}{\mddefault}{\updefault}{\color[rgb]{0,0,0}MMSE}%
}}}}
\put(18678,-19533){\makebox(0,0)[b]{\smash{{\SetFigFont{7}{8.4}{\sfdefault}{\mddefault}{\updefault}{\color[rgb]{0,0,0}estimation}%
}}}}
{\color[rgb]{0,0,0}\put(4127,-13917){\framebox(2551,1889){}}
}%
\put(5402,-12878){\makebox(0,0)[b]{\smash{{\SetFigFont{7}{8.4}{\sfdefault}{\mddefault}{\updefault}{\color[rgb]{0,0,0}Power}%
}}}}
\put(5402,-13398){\makebox(0,0)[b]{\smash{{\SetFigFont{7}{8.4}{\sfdefault}{\mddefault}{\updefault}{\color[rgb]{0,0,0}scaling}%
}}}}
{\color[rgb]{0,0,0}\put(17402,-13917){\framebox(2551,1889){}}
}%
\put(18678,-12878){\makebox(0,0)[b]{\smash{{\SetFigFont{7}{8.4}{\sfdefault}{\mddefault}{\updefault}{\color[rgb]{0,0,0}MMSE}%
}}}}
\put(18678,-13398){\makebox(0,0)[b]{\smash{{\SetFigFont{7}{8.4}{\sfdefault}{\mddefault}{\updefault}{\color[rgb]{0,0,0}estimation}%
}}}}
{\color[rgb]{0,0,0}\put(5122,-14616){\vector(-1, 0){2884}}
\put(5122,-14615){\line( 0,-1){1905}}
}%
{\color[rgb]{0,0,0}\put(6678,-12955){\vector( 1, 0){5339}}
}%
{\color[rgb]{0,0,0}\put(12478,-12969){\vector( 1, 0){4930}}
}%
{\color[rgb]{0,0,0}\put(12254,-14569){\vector( 0, 1){1375}}
}%
{\color[rgb]{0,0,0}\put(4864,-16506){\vector( 1, 0){1201}}
}%
{\color[rgb]{0,0,0}\put(2994,-17451){\framebox(1842,1890){}}
}%
{\color[rgb]{0,0,0}\put(1293,-19104){\vector( 1, 0){2835}}
}%
{\color[rgb]{0,0,0}\put(18584,-17687){\vector(-1, 0){3307}}
\put(18584,-17687){\line( 0, 1){1197}}
}%
{\color[rgb]{0,0,0}\put(19943,-19105){\vector( 1, 0){4972}}
}%
{\color[rgb]{0,0,0}\put(19943,-12955){\line( 1, 0){3318}}
\put(23261,-12962){\vector( 0,-1){3307}}
}%
{\color[rgb]{0,0,0}\put(18092,-16506){\vector( 1, 0){1910}}
}%
{\color[rgb]{0,0,0}\put(21871,-16506){\vector( 1, 0){1154}}
}%
{\color[rgb]{0,0,0}\put(23478,-16506){\vector( 1, 0){1437}}
}%
{\color[rgb]{0,0,0}\put(16269,-17451){\framebox(1843,1890){}}
}%
{\color[rgb]{0,0,0}\put(12211,-18169){\vector( 0, 1){1374}}
}%
{\color[rgb]{0,0,0}\put(15037,-16580){\vector( 0,-1){892}}
}%
{\color[rgb]{0,0,0}\put(15051,-19120){\vector( 1, 0){2342}}
\put(15051,-19115){\line( 0, 1){1145}}
}%
{\color[rgb]{0,0,0}\put(2002,-16580){\vector( 0, 1){1681}}
}%
{\color[rgb]{0,0,0}\put(2002,-12985){\vector( 1, 0){2111}}
\put(2002,-12988){\line( 0,-1){1439}}
}%
{\color[rgb]{0,0,0}\put(1199,-16573){\vector( 1, 0){1748}}
}%
{\color[rgb]{0,0,0}\put(20002,-17451){\framebox(1842,1890){}}
}%
{\color[rgb]{0,0,0}\put(6663,-19105){\line( 1, 0){2876}}
\put(9539,-19100){\vector( 0, 1){2310}}
}%
\put(1830,-18909){\makebox(0,0)[b]{\smash{{\SetFigFont{7}{8.4}{\familydefault}{\mddefault}{\updefault}{\color[rgb]{0,0,0}$S_2$}%
}}}}
\put(2521,-14474){\makebox(0,0)[b]{\smash{{\SetFigFont{7}{8.4}{\familydefault}{\mddefault}{\updefault}{\color[rgb]{0,0,0}$-$}%
}}}}
\put(24206,-16364){\makebox(0,0)[b]{\smash{{\SetFigFont{7}{8.4}{\sfdefault}{\mddefault}{\updefault}{\color[rgb]{0,0,0}$\widehat{S}_1$}%
}}}}
\put(24112,-18962){\makebox(0,0)[b]{\smash{{\SetFigFont{7}{8.4}{\sfdefault}{\mddefault}{\updefault}{\color[rgb]{0,0,0}$\widehat{S}_2$}%
}}}}
\put(8616,-12726){\makebox(0,0)[b]{\smash{{\SetFigFont{7}{8.4}{\familydefault}{\mddefault}{\updefault}{\color[rgb]{0,0,0}$(P_1)$}%
}}}}
\put(15561,-17592){\makebox(0,0)[b]{\smash{{\SetFigFont{7}{8.4}{\familydefault}{\mddefault}{\updefault}{\color[rgb]{0,0,0}$-$}%
}}}}
\put(17167,-16411){\makebox(0,0)[b]{\smash{{\SetFigFont{7}{8.4}{\sfdefault}{\mddefault}{\updefault}{\color[rgb]{0,0,0}channel}%
}}}}
\put(17167,-16931){\makebox(0,0)[b]{\smash{{\SetFigFont{7}{8.4}{\sfdefault}{\mddefault}{\updefault}{\color[rgb]{0,0,0}decoder}%
}}}}
\put(12395,-14899){\makebox(0,0)[b]{\smash{{\SetFigFont{7}{8.4}{\familydefault}{\mddefault}{\updefault}{\color[rgb]{0,0,0}$Z_{w_1}$}%
}}}}
\put(13032,-16325){\makebox(0,0)[b]{\smash{{\SetFigFont{7}{8.4}{\familydefault}{\mddefault}{\updefault}{\color[rgb]{0,0,0}$Y_{w_2}$}%
}}}}
\put(11143,-16325){\makebox(0,0)[b]{\smash{{\SetFigFont{7}{8.4}{\familydefault}{\mddefault}{\updefault}{\color[rgb]{0,0,0}$X_2$}%
}}}}
\put(11175,-12733){\makebox(0,0)[b]{\smash{{\SetFigFont{7}{8.4}{\familydefault}{\mddefault}{\updefault}{\color[rgb]{0,0,0}$X_1$}%
}}}}
\put(8861,-19718){\makebox(0,0)[b]{\smash{{\SetFigFont{7}{8.4}{\familydefault}{\mddefault}{\updefault}{\color[rgb]{0,0,0}$(P_2^{\prime})$}%
}}}}
\put(8826,-16197){\makebox(0,0)[b]{\smash{{\SetFigFont{7}{8.4}{\familydefault}{\mddefault}{\updefault}{\color[rgb]{0,0,0}$(P_2-P_2^{\prime})$}%
}}}}
\put(13075,-12724){\makebox(0,0)[b]{\smash{{\SetFigFont{7}{8.4}{\familydefault}{\mddefault}{\updefault}{\color[rgb]{0,0,0}$Y_{w_1}$}%
}}}}
\put(1783,-17098){\makebox(0,0)[b]{\smash{{\SetFigFont{7}{8.4}{\familydefault}{\mddefault}{\updefault}{\color[rgb]{0,0,0}$S_1$}%
}}}}
\end{picture}%
}\\
(b)
\end{center}
\caption{Weak-user-optimal case: (a) separation scheme showing successive
refinement (SR) and superposition coding, and (b) the hybrid digital-analog
scheme.}
\label{hybrid.fig:weak-user optimal}
\end{figure}
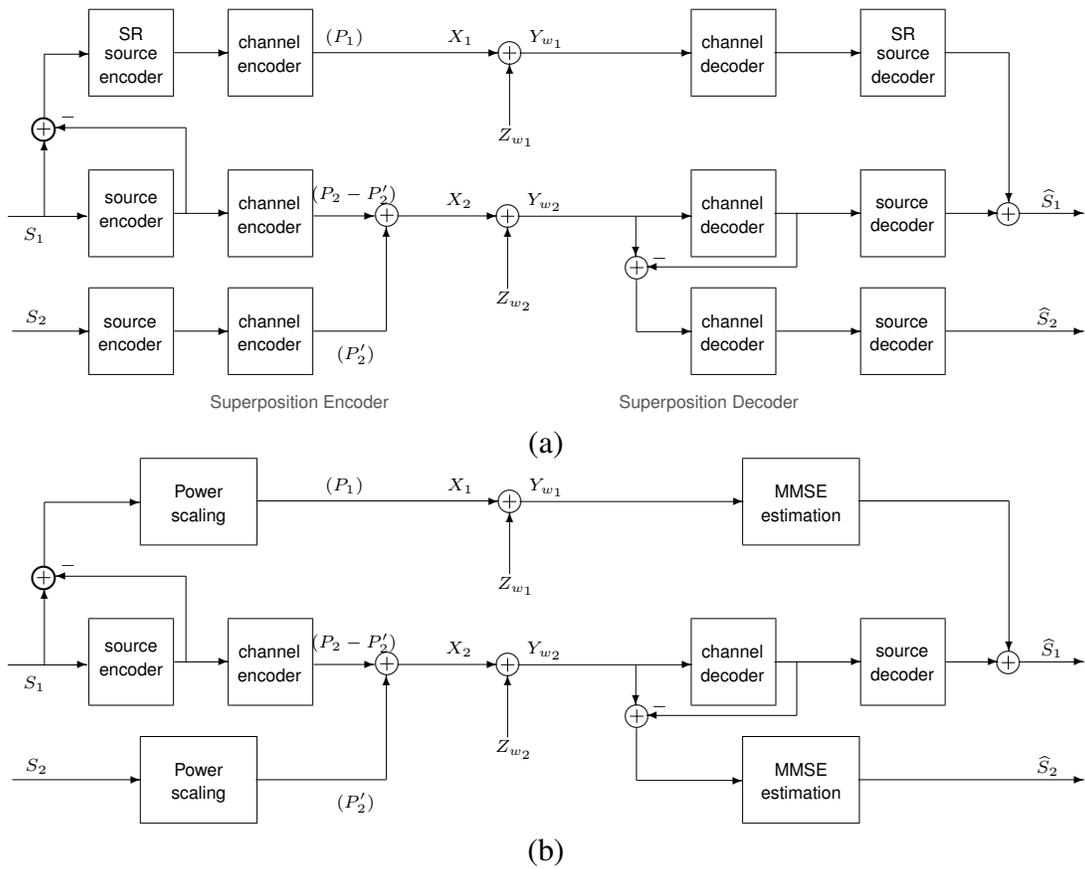

In other words, without loss of optimality, we may send the coarse
description for $S_1$ and the bit stream for $S_2$ over the second
sub-channel, and the refinement bitstream for $S_1$ over the first
sub-channel; see Fig.~\ref{hybrid.fig:weak-user optimal}(a). This further
suggests that instead of sending the refinement bitstream over the first
sub-channel, we may send uncoded the quantization error resulting from the
coarse quantization of $S_1$ appropriately scaled to satisfy the power
constraint. The input to the first sub-channel will be
$\sqrt{(P_1/D_1^\prime)}\left(S_1(i) - \widehat{S}_1^\prime(i)\right)$,
where $\widehat{S}_1^\prime(i)$ is the $i$-th sample of the coarsely
quantized version of $S_1$. It is easy to see that this satisfies the power
constraint on the first sub-channel and also results in no loss of
optimality for the weak user. The second fact is analogous to the
optimality of uncoded transmission for the point-to-point Gaussian
source-channel problem. The strong user can achieve a lower distortion on
$S_1$ because of the uncoded transmission of the quantization error. The
strong user estimates the refinement component as
$(P_1/(P_1+N_s))\sqrt{D_1^\prime/P_1}Y_1$ and adds it to the coarse
description to form its reproduction of $S_1$. We note that the resulting
distortion for the strong user on $S_1$ is ${D_1^\prime}/(1+{P_1}/{N_s})$.

The performance of the strong user can be further improved. Without losing
optimality for the weak user, we may send the coarse description of $S_1$
and the bit stream for $S_2$ using superposition coding over the second
sub-channel. In particular, we send the coarse quantization bit stream of
$S_1$ using power $P_2-P_2^\prime$ (defined below) and the bit stream for $S_2$
using power $P_2^\prime$ such that the decoder can first decode the
former bit stream assuming the latter as interference. The decoder then
cancels the interference from the bit stream for $S_1$ and decodes the bit
stream for $S_2$. Thus $P_2^\prime$ is given by
\[ \frac{1}{2}\log\left(\frac{\sigma_1^2}{D_1^\prime}\right)
 =\frac{1}{2}\log\left(1+\frac{P_2-P_2^\prime}{P_2^\prime+{N_w}}\right).\]
This also gives the relation
\[ \frac{1}{2}\log\left(\frac{\sigma_2^2}{D_2}\right)
 =\frac{1}{2}\log\left(1+\frac{P_2^\prime}{{N_w}}\right)\]
which indicates why decoding of the bit stream for $S_2$ after interference
cancellation succeeds. This scheme suggests that we may send $S_2$ uncoded
using power $P_2^\prime$ instead of sending its quantized bits. Since,
after canceling the interference from the $S_1$ bit stream, the channel to
the weak user is an additive white Gaussian noise channel, the optimality of
this scheme follows
from the optimality of uncoded transmission for point-to-point Gaussian
source-channel coding. The strong user can now reconstruct $S_2$ at a lower
distortion, $D_2(1 + P_2^\prime/N_w)/(1+P_2^\prime/N_s)$. The scheme is 
summarized in Fig.~\ref{hybrid.fig:weak-user optimal}(b). The overall
distortion achieved by the strong user is
\[
D_s=\frac{1}{2}\left(\frac{1+\frac{P_1}{N_w}}{1+\frac{P_1}{N_s}}D_1 +
\frac{1+\frac{P_2^\prime}{N_w}}{1+\frac{P_2^\prime}{N_s}}D_2\right).\]

The extension to $K=M>2$ is straightforward. Let us assume without loss of
generality that under the point-to-point optimal inverse water-filling solution
for the weak user, the first $L$ source components satisfy
$(1/2)\log(\sigma_k^2/D_k) \ge (1/2) \log(1+P_k/N_w)$. For these $L$ components
we define $D_k^\prime$ such that $(1/2)\log(D_k^\prime/D_k) = (1/2)$ $
\log(1+P_k/N_w)$. The $k$-th such component ($k\le L$) is source coded to a
distortion of $D_k^\prime$ and the resulting error is sent uncoded (scaled by
$\sqrt{P_k/D_k}$) over the $k$-th sub-channel. For sub-channels $m>L$, we define
$P_m^\prime$ as $(1/2)\log(\sigma_m^2/D_m) = (1/2) \log(1+P_m^\prime/N_w)$. The
$m$-th such component is sent uncoded over the $m$-th sub-channel scaled by
$\sqrt{P_m^\prime/\sigma_m^2}$. The rest of the power ($P_m - P_m^\prime$) for
these sub-channels $m>L$ are used to send the source coded bits from the first
$L$ components. On these sub-channels, the decoders first decode these bits,
cancel the interference caused by them, and then estimate the source
components. The first $L$ source components are estimated directly from the
corresponding sub-channel outputs. Thus, without compromising the quality of
reproduction for the weak user, the strong user achieves a lower distortion.
\[ D_s=\frac{1}{K}\left(
\sum_{k=1}^L \frac{1+\frac{P_k}{N_w}}{1+\frac{P_k}{N_s}}D_k +
\sum_{k=L+1}^K \frac{1+\frac{P_k^\prime}{N_w}}{1+\frac{P_k^\prime}{N_s}}D_k
\right)
 < \frac{1}{K}\sum_{k=1}^K D_k. \]

This scheme directly extends to the $K\neq M$ case. When $K<M$ (bandwidth
expansion), the extra sub-channels will be used to send additional coded bits.
For the case of $K>M$ (bandwidth contraction), only at most $M$ source
components can be sent (wholly or partially) uncoded. In summary, we have
the following
\begin{thm}
For the source-channel problem in Section~\ref{hybrid.sec:problemstatement}
$(D_s,D_w^\ast)$ is achievable, where $D_w^\ast$ is given by
(\ref{hybrid.eq:reversewaterfill}) and $D_s$ is as defined below.
\[D_s =\frac{1}{K}\left(
\sum_{k=1}^L \frac{1+\frac{P_k}{N_w}}{1+\frac{P_k}{N_s}}D_k +
\sum_{k=L+1}^{K^\prime}
\frac{1+\frac{P_k^\prime}{N_w}}{1+\frac{P_k^\prime}{N_s}}D_k +
\sum_{k=K^\prime+1}^{K} D_k\right),\]
where $D_k$'s are given by (\ref{hybrid.eq:reversewaterfill}), $P_k=P$, 
\[L=\min\left\{\left|\left\{k:\frac{\sigma_k^2}{D_k}\ge 1 +
\frac{P_k}{N_w}\right\}\right|,M\right\},\]
\[K^\prime=\min\left\{\left|\left\{k:\mu\le\sigma_k^2\right\}\right|,K\right\},\]
and the $P_k^\prime$'s are defined by
\[\frac{\sigma_k^2}{D_k} = 1+\frac{P_k^\prime}{N_w}, \;k=L+1,\ldots,K^\prime.\]
 
\end{thm}

\subsection{Strong-user-optimal case}

If the point-to-point separation approach is used for providing optimal
fidelity to the strong user, since the rate of transmission is greater than
the channel capacity of the weak user, the weak user will not be able to
get any useful information. However, in this subsection we show that we can
provide useful information to the weak user without compromising the strong
user's performance. As will become clear, in this case, the weak user's
receiver will only involve scaling the signals received on different
sub-channels, {\em i.e.}, it will be an analog receiver. An extension of
the scheme in this subsection was shown in~\cite{prabhakaranprupgrade08} to
obtain the entire optimal distortion trade-off region for broadcasting a
parallel Gaussian source over a parallel Gaussian broadcast channel to two
receivers when one of the receivers is restricted to be a linear filter
(but without no assumptions on the relative strengths of the channels to
the receivers). The linear filter receiver is a model for a legacy analog
receiver in a transitionary broadcast system which supports digital and
analog receivers.

Let the point-to-point optimal reverse water-filling solution for the
strong user produce a total distortion $D_s^\ast$ from a distortion
allocation $D_k, \; k=1,2,\ldots,K$ according to
(\ref{hybrid.eq:reversewaterfill}), where $\mu$ is now chosen so that the
total rate equals the capacity $C_s=(1/2)\sum_{m=1}^M\log(1+P_m/N_s),\;
P_m=P$ of the strong user's channel.

\begin{figure}[htb]
\begin{center}
\scalebox{0.95}{\setlength{\unitlength}{1184sp}%
\begingroup\makeatletter\ifx\SetFigFont\undefined%
\gdef\SetFigFont#1#2#3#4#5{%
  \reset@font\fontsize{#1}{#2pt}%
  \fontfamily{#3}\fontseries{#4}\fontshape{#5}%
  \selectfont}%
\fi\endgroup%
\begin{picture}(23735,9195)(393,-22271)
{\color[rgb]{0,0,0}\thicklines
\put(1166,-15727){\circle{472}}
}%
{\color[rgb]{0,0,0}\put(1166,-15577){\line( 0,-1){301}}
}%
{\color[rgb]{0,0,0}\put(1017,-15727){\line( 1, 0){298}}
}%
{\color[rgb]{0,0,0}\put(8725,-20168){\circle{472}}
}%
{\color[rgb]{0,0,0}\put(8725,-20018){\line( 0,-1){301}}
}%
{\color[rgb]{0,0,0}\put(8576,-20168){\line( 1, 0){298}}
}%
{\color[rgb]{0,0,0}\put(2174,-14987){\framebox(1842,1889){}}
}%
\put(3071,-13948){\makebox(0,0)[b]{\smash{{\SetFigFont{7}{8.4}{\sfdefault}{\mddefault}{\updefault}{\color[rgb]{0,0,0}source}%
}}}}
\put(3071,-14468){\makebox(0,0)[b]{\smash{{\SetFigFont{7}{8.4}{\sfdefault}{\mddefault}{\updefault}{\color[rgb]{0,0,0}encoder}%
}}}}
{\color[rgb]{0,0,0}\put(5245,-14987){\framebox(1842,1889){}}
}%
\put(6142,-13948){\makebox(0,0)[b]{\smash{{\SetFigFont{7}{8.4}{\sfdefault}{\mddefault}{\updefault}{\color[rgb]{0,0,0}channel}%
}}}}
\put(6142,-14468){\makebox(0,0)[b]{\smash{{\SetFigFont{7}{8.4}{\sfdefault}{\mddefault}{\updefault}{\color[rgb]{0,0,0}encoder}%
}}}}
{\color[rgb]{0,0,0}\put(11434,-14049){\circle{472}}
}%
{\color[rgb]{0,0,0}\put(11434,-13899){\line( 0,-1){300}}
}%
{\color[rgb]{0,0,0}\put(11284,-14049){\line( 1, 0){299}}
}%
{\color[rgb]{0,0,0}\put(11386,-20190){\circle{472}}
}%
{\color[rgb]{0,0,0}\put(11386,-20040){\line( 0,-1){301}}
}%
{\color[rgb]{0,0,0}\put(11237,-20190){\line( 1, 0){298}}
}%
{\color[rgb]{0,0,0}\put(11386,-21790){\vector( 0, 1){1374}}
}%
{\color[rgb]{0,0,0}\put(5245,-21129){\framebox(1842,1890){}}
}%
\put(6142,-20090){\makebox(0,0)[b]{\smash{{\SetFigFont{7}{8.4}{\sfdefault}{\mddefault}{\updefault}{\color[rgb]{0,0,0}channel}%
}}}}
\put(6142,-20609){\makebox(0,0)[b]{\smash{{\SetFigFont{7}{8.4}{\sfdefault}{\mddefault}{\updefault}{\color[rgb]{0,0,0}encoder}%
}}}}
{\color[rgb]{0,0,0}\put(5245,-18531){\framebox(1842,1890){}}
}%
\put(6142,-17491){\makebox(0,0)[b]{\smash{{\SetFigFont{7}{8.4}{\sfdefault}{\mddefault}{\updefault}{\color[rgb]{0,0,0}channel}%
}}}}
\put(6142,-18011){\makebox(0,0)[b]{\smash{{\SetFigFont{7}{8.4}{\sfdefault}{\mddefault}{\updefault}{\color[rgb]{0,0,0}encoder}%
}}}}
{\color[rgb]{0,0,0}\put(2174,-21129){\framebox(1842,1890){}}
}%
\put(3071,-20090){\makebox(0,0)[b]{\smash{{\SetFigFont{7}{8.4}{\sfdefault}{\mddefault}{\updefault}{\color[rgb]{0,0,0}source}%
}}}}
\put(3071,-20609){\makebox(0,0)[b]{\smash{{\SetFigFont{7}{8.4}{\sfdefault}{\mddefault}{\updefault}{\color[rgb]{0,0,0}encoder}%
}}}}
{\color[rgb]{0,0,0}\put(14237,-18784){\circle{472}}
}%
{\color[rgb]{0,0,0}\put(14237,-18633){\line( 0,-1){301}}
}%
{\color[rgb]{0,0,0}\put(14088,-18784){\line( 1, 0){298}}
}%
{\color[rgb]{0,0,0}\put(15449,-21129){\framebox(1843,1890){}}
}%
\put(16347,-20090){\makebox(0,0)[b]{\smash{{\SetFigFont{7}{8.4}{\sfdefault}{\mddefault}{\updefault}{\color[rgb]{0,0,0}channel}%
}}}}
\put(16347,-20609){\makebox(0,0)[b]{\smash{{\SetFigFont{7}{8.4}{\sfdefault}{\mddefault}{\updefault}{\color[rgb]{0,0,0}decoder}%
}}}}
{\color[rgb]{0,0,0}\put(19182,-21129){\framebox(1842,1890){}}
}%
\put(20079,-20090){\makebox(0,0)[b]{\smash{{\SetFigFont{7}{8.4}{\sfdefault}{\mddefault}{\updefault}{\color[rgb]{0,0,0}source}%
}}}}
\put(20079,-20609){\makebox(0,0)[b]{\smash{{\SetFigFont{7}{8.4}{\sfdefault}{\mddefault}{\updefault}{\color[rgb]{0,0,0}decoder}%
}}}}
{\color[rgb]{0,0,0}\put(22426,-17602){\circle{472}}
}%
{\color[rgb]{0,0,0}\put(22426,-17452){\line( 0,-1){301}}
}%
{\color[rgb]{0,0,0}\put(22277,-17602){\line( 1, 0){298}}
}%
\put(10365,-13804){\makebox(0,0)[b]{\smash{{\SetFigFont{7}{8.4}{\familydefault}{\mddefault}{\updefault}{\color[rgb]{0,0,0}$X_1$}%
}}}}
\put(10318,-19946){\makebox(0,0)[b]{\smash{{\SetFigFont{7}{8.4}{\familydefault}{\mddefault}{\updefault}{\color[rgb]{0,0,0}$X_2$}%
}}}}
\put(20079,-13948){\makebox(0,0)[b]{\smash{{\SetFigFont{7}{8.4}{\sfdefault}{\mddefault}{\updefault}{\color[rgb]{0,0,0}source}%
}}}}
\put(20079,-14468){\makebox(0,0)[b]{\smash{{\SetFigFont{7}{8.4}{\sfdefault}{\mddefault}{\updefault}{\color[rgb]{0,0,0}decoder}%
}}}}
{\color[rgb]{0,0,0}\put(15449,-14987){\framebox(1843,1889){}}
}%
\put(16347,-13948){\makebox(0,0)[b]{\smash{{\SetFigFont{7}{8.4}{\sfdefault}{\mddefault}{\updefault}{\color[rgb]{0,0,0}channel}%
}}}}
\put(16347,-14468){\makebox(0,0)[b]{\smash{{\SetFigFont{7}{8.4}{\sfdefault}{\mddefault}{\updefault}{\color[rgb]{0,0,0}decoder}%
}}}}
{\color[rgb]{0,0,0}\put(19182,-14987){\framebox(1842,1889){}}
}%
{\color[rgb]{0,0,0}\put(17272,-14042){\vector( 1, 0){1910}}
}%
\put(12255,-13804){\makebox(0,0)[b]{\smash{{\SetFigFont{7}{8.4}{\familydefault}{\mddefault}{\updefault}{\color[rgb]{0,0,0}$Y_{s_1}$}%
}}}}
\put(12207,-19946){\makebox(0,0)[b]{\smash{{\SetFigFont{7}{8.4}{\familydefault}{\mddefault}{\updefault}{\color[rgb]{0,0,0}$Y_{s_2}$}%
}}}}
{\color[rgb]{0,0,0}\put(1182,-14068){\vector( 0,-1){1439}}
}%
{\color[rgb]{0,0,0}\put(4489,-15696){\vector(-1, 0){3071}}
\put(4489,-15696){\line( 0, 1){1606}}
}%
{\color[rgb]{0,0,0}\put(426,-14042){\vector( 1, 0){1748}}
}%
{\color[rgb]{0,0,0}\put(4044,-14042){\vector( 1, 0){1201}}
}%
{\color[rgb]{0,0,0}\put(7087,-14042){\vector( 1, 0){4110}}
}%
{\color[rgb]{0,0,0}\put(8957,-20184){\vector( 1, 0){2193}}
}%
{\color[rgb]{0,0,0}\put(11658,-14049){\vector( 1, 0){3791}}
}%
{\color[rgb]{0,0,0}\put(11434,-15649){\vector( 0, 1){1375}}
}%
{\color[rgb]{0,0,0}\put(11611,-20190){\vector( 1, 0){3838}}
}%
{\color[rgb]{0,0,0}\put(7114,-20184){\vector( 1, 0){1343}}
}%
{\color[rgb]{0,0,0}\put(7087,-17586){\line( 1, 0){1654}}
\put(8741,-17586){\vector( 0,-1){2362}}
}%
{\color[rgb]{0,0,0}}%
{\color[rgb]{0,0,0}\put(4044,-17586){\vector( 1, 0){1201}}
}%
{\color[rgb]{0,0,0}\put(4044,-20184){\vector( 1, 0){1201}}
}%
{\color[rgb]{0,0,0}\put(1182,-15979){\line( 0,-1){1607}}
\put(1182,-17586){\vector( 1, 0){992}}
}%
{\color[rgb]{0,0,0}\put(2174,-18531){\framebox(1842,1890){}}
}%
{\color[rgb]{0,0,0}\put(473,-20184){\vector( 1, 0){1701}}
}%
{\color[rgb]{0,0,0}\put(17319,-20184){\vector( 1, 0){1863}}
}%
{\color[rgb]{0,0,0}\put(14221,-20184){\vector( 0, 1){1134}}
}%
{\color[rgb]{0,0,0}\put(14221,-18531){\line( 0, 1){945}}
\put(14221,-17586){\vector( 1, 0){1228}}
}%
{\color[rgb]{0,0,0}\put(17764,-18767){\vector(-1, 0){3307}}
\put(17764,-18767){\line( 0,-1){1417}}
}%
{\color[rgb]{0,0,0}}%
{\color[rgb]{0,0,0}\put(21051,-20184){\vector( 1, 0){3044}}
}%
{\color[rgb]{0,0,0}\put(21024,-14042){\line( 1, 0){1417}}
\put(22441,-14042){\vector( 0,-1){3307}}
}%
{\color[rgb]{0,0,0}\put(19182,-18531){\framebox(1842,1890){}}
}%
{\color[rgb]{0,0,0}\put(17272,-17586){\vector( 1, 0){1910}}
}%
{\color[rgb]{0,0,0}\put(21051,-17586){\vector( 1, 0){1154}}
}%
{\color[rgb]{0,0,0}\put(22658,-17586){\vector( 1, 0){1437}}
}%
{\color[rgb]{0,0,0}\put(15449,-18531){\framebox(1843,1890){}}
}%
\put(1010,-19989){\makebox(0,0)[b]{\smash{{\SetFigFont{7}{8.4}{\familydefault}{\mddefault}{\updefault}{\color[rgb]{0,0,0}$S_2$}%
}}}}
\put(1701,-15554){\makebox(0,0)[b]{\smash{{\SetFigFont{7}{8.4}{\familydefault}{\mddefault}{\updefault}{\color[rgb]{0,0,0}$-$}%
}}}}
\put(1010,-13847){\makebox(0,0)[b]{\smash{{\SetFigFont{7}{8.4}{\familydefault}{\mddefault}{\updefault}{\color[rgb]{0,0,0}$S_1$}%
}}}}
\put(11528,-22121){\makebox(0,0)[b]{\smash{{\SetFigFont{7}{8.4}{\familydefault}{\mddefault}{\updefault}{\color[rgb]{0,0,0}$Z_{s_2}$}%
}}}}
\put(3071,-17680){\makebox(0,0)[b]{\smash{{\SetFigFont{7}{8.4}{\sfdefault}{\mddefault}{\updefault}{\color[rgb]{0,0,0}source}%
}}}}
\put(3071,-18200){\makebox(0,0)[b]{\smash{{\SetFigFont{7}{8.4}{\sfdefault}{\mddefault}{\updefault}{\color[rgb]{0,0,0}encoder}%
}}}}
\put(3071,-17255){\makebox(0,0)[b]{\smash{{\SetFigFont{7}{8.4}{\sfdefault}{\mddefault}{\updefault}{\color[rgb]{0,0,0}SR}%
}}}}
\put(20079,-17255){\makebox(0,0)[b]{\smash{{\SetFigFont{7}{8.4}{\sfdefault}{\mddefault}{\updefault}{\color[rgb]{0,0,0}SR}%
}}}}
\put(20079,-17680){\makebox(0,0)[b]{\smash{{\SetFigFont{7}{8.4}{\sfdefault}{\mddefault}{\updefault}{\color[rgb]{0,0,0}source}%
}}}}
\put(20079,-18200){\makebox(0,0)[b]{\smash{{\SetFigFont{7}{8.4}{\sfdefault}{\mddefault}{\updefault}{\color[rgb]{0,0,0}decoder}%
}}}}
\put(23386,-17444){\makebox(0,0)[b]{\smash{{\SetFigFont{7}{8.4}{\sfdefault}{\mddefault}{\updefault}{\color[rgb]{0,0,0}$\widehat{S}_1$}%
}}}}
\put(23292,-20042){\makebox(0,0)[b]{\smash{{\SetFigFont{7}{8.4}{\sfdefault}{\mddefault}{\updefault}{\color[rgb]{0,0,0}$\widehat{S}_2$}%
}}}}
\put(7796,-13806){\makebox(0,0)[b]{\smash{{\SetFigFont{7}{8.4}{\familydefault}{\mddefault}{\updefault}{\color[rgb]{0,0,0}$(P_1)$}%
}}}}
\put(7796,-17397){\makebox(0,0)[b]{\smash{{\SetFigFont{7}{8.4}{\familydefault}{\mddefault}{\updefault}{\color[rgb]{0,0,0}$(P_2^{\prime\prime})$}%
}}}}
\put(14741,-18672){\makebox(0,0)[b]{\smash{{\SetFigFont{7}{8.4}{\familydefault}{\mddefault}{\updefault}{\color[rgb]{0,0,0}$-$}%
}}}}
\put(15827,-21885){\makebox(0,0)[b]{\smash{{\SetFigFont{7}{8.4}{\sfdefault}{\mddefault}{\updefault}{\color[rgb]{0.333,0.333,0.333}Superposition Decoder}%
}}}}
\put(6804,-21885){\makebox(0,0)[b]{\smash{{\SetFigFont{7}{8.4}{\sfdefault}{\mddefault}{\updefault}{\color[rgb]{0.333,0.333,0.333}Superposition Encoder}%
}}}}
\put(16347,-17491){\makebox(0,0)[b]{\smash{{\SetFigFont{7}{8.4}{\sfdefault}{\mddefault}{\updefault}{\color[rgb]{0,0,0}channel}%
}}}}
\put(16347,-18011){\makebox(0,0)[b]{\smash{{\SetFigFont{7}{8.4}{\sfdefault}{\mddefault}{\updefault}{\color[rgb]{0,0,0}decoder}%
}}}}
\put(11575,-15979){\makebox(0,0)[b]{\smash{{\SetFigFont{7}{8.4}{\familydefault}{\mddefault}{\updefault}{\color[rgb]{0,0,0}$Z_{s_1}$}%
}}}}
\put(8041,-20798){\makebox(0,0)[b]{\smash{{\SetFigFont{7}{8.4}{\familydefault}{\mddefault}{\updefault}{\color[rgb]{0,0,0}$(P_2-P_2^{\prime\prime})$}%
}}}}
\end{picture}%
}\\
(a)\\
\scalebox{0.95}{\setlength{\unitlength}{1184sp}%
\begingroup\makeatletter\ifx\SetFigFont\undefined%
\gdef\SetFigFont#1#2#3#4#5{%
  \reset@font\fontsize{#1}{#2pt}%
  \fontfamily{#3}\fontseries{#4}\fontshape{#5}%
  \selectfont}%
\fi\endgroup%
\begin{picture}(23735,9193)(94,-11894)
{\color[rgb]{0,0,0}\thicklines
\put(8426,-9791){\circle{472}}
}%
{\color[rgb]{0,0,0}\put(8426,-9641){\line( 0,-1){301}}
}%
{\color[rgb]{0,0,0}\put(8277,-9791){\line( 1, 0){298}}
}%
{\color[rgb]{0,0,0}\put(11135,-3672){\circle{472}}
}%
{\color[rgb]{0,0,0}\put(11135,-3522){\line( 0,-1){300}}
}%
{\color[rgb]{0,0,0}\put(10985,-3672){\line( 1, 0){299}}
}%
{\color[rgb]{0,0,0}\put(11087,-9813){\circle{472}}
}%
{\color[rgb]{0,0,0}\put(11087,-9663){\line( 0,-1){301}}
}%
{\color[rgb]{0,0,0}\put(10938,-9813){\line( 1, 0){298}}
}%
{\color[rgb]{0,0,0}\put(11087,-11413){\vector( 0, 1){1374}}
}%
{\color[rgb]{0,0,0}\put(4946,-10752){\framebox(1842,1890){}}
}%
\put(5843,-9713){\makebox(0,0)[b]{\smash{{\SetFigFont{7}{8.4}{\sfdefault}{\mddefault}{\updefault}{\color[rgb]{0,0,0}channel}%
}}}}
\put(5843,-10232){\makebox(0,0)[b]{\smash{{\SetFigFont{7}{8.4}{\sfdefault}{\mddefault}{\updefault}{\color[rgb]{0,0,0}encoder}%
}}}}
{\color[rgb]{0,0,0}\put(1875,-10752){\framebox(1842,1890){}}
}%
\put(2772,-9713){\makebox(0,0)[b]{\smash{{\SetFigFont{7}{8.4}{\sfdefault}{\mddefault}{\updefault}{\color[rgb]{0,0,0}source}%
}}}}
\put(2772,-10232){\makebox(0,0)[b]{\smash{{\SetFigFont{7}{8.4}{\sfdefault}{\mddefault}{\updefault}{\color[rgb]{0,0,0}encoder}%
}}}}
{\color[rgb]{0,0,0}\put(15150,-10752){\framebox(1843,1890){}}
}%
\put(16048,-9713){\makebox(0,0)[b]{\smash{{\SetFigFont{7}{8.4}{\sfdefault}{\mddefault}{\updefault}{\color[rgb]{0,0,0}channel}%
}}}}
\put(16048,-10232){\makebox(0,0)[b]{\smash{{\SetFigFont{7}{8.4}{\sfdefault}{\mddefault}{\updefault}{\color[rgb]{0,0,0}decoder}%
}}}}
{\color[rgb]{0,0,0}\put(18883,-10752){\framebox(1842,1890){}}
}%
\put(19780,-9713){\makebox(0,0)[b]{\smash{{\SetFigFont{7}{8.4}{\sfdefault}{\mddefault}{\updefault}{\color[rgb]{0,0,0}source}%
}}}}
\put(19780,-10232){\makebox(0,0)[b]{\smash{{\SetFigFont{7}{8.4}{\sfdefault}{\mddefault}{\updefault}{\color[rgb]{0,0,0}decoder}%
}}}}
\put(10066,-3427){\makebox(0,0)[b]{\smash{{\SetFigFont{7}{8.4}{\familydefault}{\mddefault}{\updefault}{\color[rgb]{0,0,0}$X_1$}%
}}}}
\put(10019,-9569){\makebox(0,0)[b]{\smash{{\SetFigFont{7}{8.4}{\familydefault}{\mddefault}{\updefault}{\color[rgb]{0,0,0}$X_2$}%
}}}}
\put(11956,-3427){\makebox(0,0)[b]{\smash{{\SetFigFont{7}{8.4}{\familydefault}{\mddefault}{\updefault}{\color[rgb]{0,0,0}$Y_{s_1}$}%
}}}}
\put(11908,-9569){\makebox(0,0)[b]{\smash{{\SetFigFont{7}{8.4}{\familydefault}{\mddefault}{\updefault}{\color[rgb]{0,0,0}$Y_{s_2}$}%
}}}}
{\color[rgb]{0,0,0}\put(1871,-4612){\framebox(1842,1889){}}
}%
\put(2768,-3573){\makebox(0,0)[b]{\smash{{\SetFigFont{7}{8.4}{\sfdefault}{\mddefault}{\updefault}{\color[rgb]{0,0,0}source}%
}}}}
\put(2768,-4093){\makebox(0,0)[b]{\smash{{\SetFigFont{7}{8.4}{\sfdefault}{\mddefault}{\updefault}{\color[rgb]{0,0,0}encoder}%
}}}}
{\color[rgb]{0,0,0}\put(4942,-4612){\framebox(1842,1889){}}
}%
\put(5839,-3573){\makebox(0,0)[b]{\smash{{\SetFigFont{7}{8.4}{\sfdefault}{\mddefault}{\updefault}{\color[rgb]{0,0,0}channel}%
}}}}
\put(5839,-4093){\makebox(0,0)[b]{\smash{{\SetFigFont{7}{8.4}{\sfdefault}{\mddefault}{\updefault}{\color[rgb]{0,0,0}encoder}%
}}}}
\put(19776,-3573){\makebox(0,0)[b]{\smash{{\SetFigFont{7}{8.4}{\sfdefault}{\mddefault}{\updefault}{\color[rgb]{0,0,0}source}%
}}}}
\put(19776,-4093){\makebox(0,0)[b]{\smash{{\SetFigFont{7}{8.4}{\sfdefault}{\mddefault}{\updefault}{\color[rgb]{0,0,0}decoder}%
}}}}
{\color[rgb]{0,0,0}\put(15146,-4612){\framebox(1843,1889){}}
}%
\put(16044,-3573){\makebox(0,0)[b]{\smash{{\SetFigFont{7}{8.4}{\sfdefault}{\mddefault}{\updefault}{\color[rgb]{0,0,0}channel}%
}}}}
\put(16044,-4093){\makebox(0,0)[b]{\smash{{\SetFigFont{7}{8.4}{\sfdefault}{\mddefault}{\updefault}{\color[rgb]{0,0,0}decoder}%
}}}}
{\color[rgb]{0,0,0}\put(18879,-4612){\framebox(1842,1889){}}
}%
{\color[rgb]{0,0,0}\put(16969,-3667){\vector( 1, 0){1910}}
}%
{\color[rgb]{0,0,0}\put(3741,-3667){\vector( 1, 0){1201}}
}%
{\color[rgb]{0,0,0}\put(6784,-3667){\vector( 1, 0){4110}}
}%
{\color[rgb]{0,0,0}\put(11355,-3674){\vector( 1, 0){3791}}
}%
{\color[rgb]{0,0,0}\put(127,-3665){\vector( 1, 0){1756}}
}%
{\color[rgb]{0,0,0}\put(8658,-9807){\vector( 1, 0){2193}}
}%
{\color[rgb]{0,0,0}\put(11135,-5272){\vector( 0, 1){1375}}
}%
{\color[rgb]{0,0,0}\put(11312,-9813){\vector( 1, 0){3838}}
}%
{\color[rgb]{0,0,0}\put(6815,-9807){\vector( 1, 0){1343}}
}%
{\color[rgb]{0,0,0}\put(6788,-7209){\line( 1, 0){1654}}
\put(8442,-7209){\vector( 0,-1){2362}}
}%
{\color[rgb]{0,0,0}}%
{\color[rgb]{0,0,0}\put(3745,-7209){\vector( 1, 0){1201}}
}%
{\color[rgb]{0,0,0}\put(3745,-9807){\vector( 1, 0){1201}}
}%
{\color[rgb]{0,0,0}\put(883,-3665){\line( 0,-1){3544}}
\put(883,-7209){\vector( 1, 0){992}}
}%
{\color[rgb]{0,0,0}\put(1875,-8154){\framebox(1842,1890){}}
}%
{\color[rgb]{0,0,0}\put(174,-9807){\vector( 1, 0){1701}}
}%
{\color[rgb]{0,0,0}\put(17020,-9807){\vector( 1, 0){1863}}
}%
{\color[rgb]{0,0,0}\put(13922,-9807){\line( 0, 1){2598}}
\put(13922,-7209){\vector( 1, 0){1228}}
}%
{\color[rgb]{0,0,0}}%
{\color[rgb]{0,0,0}\put(20752,-9807){\vector( 1, 0){3044}}
}%
{\color[rgb]{0,0,0}\put(18883,-8154){\framebox(1842,1890){}}
}%
{\color[rgb]{0,0,0}\put(16973,-7209){\vector( 1, 0){1910}}
}%
{\color[rgb]{0,0,0}\put(20725,-7209){\vector( 1, 0){3071}}
}%
{\color[rgb]{0,0,0}\put(20708,-3665){\line( 1, 0){1434}}
\put(22142,-3665){\line( 0,-1){1418}}
\put(22142,-5083){\line(-1, 0){2362}}
\put(19780,-5083){\vector( 0,-1){1181}}
}%
{\color[rgb]{0,0,0}\put(5843,-8673){\vector( 0, 1){519}}
\put(5843,-8673){\line( 1, 0){1418}}
\put(7261,-8673){\line( 0,-1){1134}}
}%
{\color[rgb]{0,0,0}\put(15150,-8154){\framebox(1843,1890){}}
}%
{\color[rgb]{0,0,0}\put(4946,-8154){\framebox(1842,1890){}}
}%
\put(711,-9612){\makebox(0,0)[b]{\smash{{\SetFigFont{7}{8.4}{\familydefault}{\mddefault}{\updefault}{\color[rgb]{0,0,0}$S_2$}%
}}}}
\put(711,-3470){\makebox(0,0)[b]{\smash{{\SetFigFont{7}{8.4}{\familydefault}{\mddefault}{\updefault}{\color[rgb]{0,0,0}$S_1$}%
}}}}
\put(11276,-5602){\makebox(0,0)[b]{\smash{{\SetFigFont{7}{8.4}{\familydefault}{\mddefault}{\updefault}{\color[rgb]{0,0,0}$Z_{s_1}$}%
}}}}
\put(11229,-11744){\makebox(0,0)[b]{\smash{{\SetFigFont{7}{8.4}{\familydefault}{\mddefault}{\updefault}{\color[rgb]{0,0,0}$Z_{s_2}$}%
}}}}
\put(2772,-7303){\makebox(0,0)[b]{\smash{{\SetFigFont{7}{8.4}{\sfdefault}{\mddefault}{\updefault}{\color[rgb]{0,0,0}source}%
}}}}
\put(2772,-7823){\makebox(0,0)[b]{\smash{{\SetFigFont{7}{8.4}{\sfdefault}{\mddefault}{\updefault}{\color[rgb]{0,0,0}encoder}%
}}}}
\put(19780,-7303){\makebox(0,0)[b]{\smash{{\SetFigFont{7}{8.4}{\sfdefault}{\mddefault}{\updefault}{\color[rgb]{0,0,0}source}%
}}}}
\put(19780,-7823){\makebox(0,0)[b]{\smash{{\SetFigFont{7}{8.4}{\sfdefault}{\mddefault}{\updefault}{\color[rgb]{0,0,0}decoder}%
}}}}
\put(23087,-7067){\makebox(0,0)[b]{\smash{{\SetFigFont{7}{8.4}{\sfdefault}{\mddefault}{\updefault}{\color[rgb]{0,0,0}$\widehat{S}_1$}%
}}}}
\put(22993,-9665){\makebox(0,0)[b]{\smash{{\SetFigFont{7}{8.4}{\sfdefault}{\mddefault}{\updefault}{\color[rgb]{0,0,0}$\widehat{S}_2$}%
}}}}
\put(7497,-7020){\makebox(0,0)[b]{\smash{{\SetFigFont{7}{8.4}{\familydefault}{\mddefault}{\updefault}{\color[rgb]{0,0,0}$(P_2^{\prime\prime})$}%
}}}}
\put(2772,-6878){\makebox(0,0)[b]{\smash{{\SetFigFont{7}{8.4}{\sfdefault}{\mddefault}{\updefault}{\color[rgb]{0,0,0}WZ}%
}}}}
\put(19780,-6878){\makebox(0,0)[b]{\smash{{\SetFigFont{7}{8.4}{\sfdefault}{\mddefault}{\updefault}{\color[rgb]{0,0,0}WZ}%
}}}}
\put(6505,-11508){\makebox(0,0)[b]{\smash{{\SetFigFont{7}{8.4}{\sfdefault}{\mddefault}{\updefault}{\color[rgb]{0.333,0.333,0.333}Dirty-Paper Encoder}%
}}}}
\put(15528,-11508){\makebox(0,0)[b]{\smash{{\SetFigFont{7}{8.4}{\sfdefault}{\mddefault}{\updefault}{\color[rgb]{0.333,0.333,0.333}Dirty-Paper Decoder}%
}}}}
\put(16048,-7634){\makebox(0,0)[b]{\smash{{\SetFigFont{7}{8.4}{\sfdefault}{\mddefault}{\updefault}{\color[rgb]{0,0,0}decoder}%
}}}}
\put(16048,-7114){\makebox(0,0)[b]{\smash{{\SetFigFont{7}{8.4}{\sfdefault}{\mddefault}{\updefault}{\color[rgb]{0,0,0}dirty-paper}%
}}}}
\put(5843,-7634){\makebox(0,0)[b]{\smash{{\SetFigFont{7}{8.4}{\sfdefault}{\mddefault}{\updefault}{\color[rgb]{0,0,0}encoder}%
}}}}
\put(5843,-7114){\makebox(0,0)[b]{\smash{{\SetFigFont{7}{8.4}{\sfdefault}{\mddefault}{\updefault}{\color[rgb]{0,0,0}dirty-paper}%
}}}}
\put(7645,-10421){\makebox(0,0)[b]{\smash{{\SetFigFont{7}{8.4}{\familydefault}{\mddefault}{\updefault}{\color[rgb]{0,0,0}$(P_2-P_2^{\prime\prime})$}%
}}}}
\end{picture}%
}\\
(b)\\
\scalebox{0.95}{\setlength{\unitlength}{1184sp}%
\begingroup\makeatletter\ifx\SetFigFont\undefined%
\gdef\SetFigFont#1#2#3#4#5{%
  \reset@font\fontsize{#1}{#2pt}%
  \fontfamily{#3}\fontseries{#4}\fontshape{#5}%
  \selectfont}%
\fi\endgroup%
\begin{picture}(23735,9195)(393,-22271)
{\color[rgb]{0,0,0}\thicklines
\put(8725,-20168){\circle{472}}
}%
{\color[rgb]{0,0,0}\put(8725,-20018){\line( 0,-1){301}}
}%
{\color[rgb]{0,0,0}\put(8576,-20168){\line( 1, 0){298}}
}%
{\color[rgb]{0,0,0}\put(11434,-14049){\circle{472}}
}%
{\color[rgb]{0,0,0}\put(11434,-13899){\line( 0,-1){300}}
}%
{\color[rgb]{0,0,0}\put(11284,-14049){\line( 1, 0){299}}
}%
{\color[rgb]{0,0,0}\put(11386,-20190){\circle{472}}
}%
{\color[rgb]{0,0,0}\put(11386,-20040){\line( 0,-1){301}}
}%
{\color[rgb]{0,0,0}\put(11237,-20190){\line( 1, 0){298}}
}%
{\color[rgb]{0,0,0}\put(11386,-21790){\vector( 0, 1){1374}}
}%
\put(10365,-13804){\makebox(0,0)[b]{\smash{{\SetFigFont{7}{8.4}{\familydefault}{\mddefault}{\updefault}{\color[rgb]{0,0,0}$X_1$}%
}}}}
\put(10318,-19946){\makebox(0,0)[b]{\smash{{\SetFigFont{7}{8.4}{\familydefault}{\mddefault}{\updefault}{\color[rgb]{0,0,0}$X_2$}%
}}}}
\put(12255,-13804){\makebox(0,0)[b]{\smash{{\SetFigFont{7}{8.4}{\familydefault}{\mddefault}{\updefault}{\color[rgb]{0,0,0}$Y_{s_1}$}%
}}}}
\put(12207,-19946){\makebox(0,0)[b]{\smash{{\SetFigFont{7}{8.4}{\familydefault}{\mddefault}{\updefault}{\color[rgb]{0,0,0}$Y_{s_2}$}%
}}}}
{\color[rgb]{0,0,0}\put(3308,-14987){\framebox(2551,1889){}}
}%
\put(4583,-13948){\makebox(0,0)[b]{\smash{{\SetFigFont{7}{8.4}{\sfdefault}{\mddefault}{\updefault}{\color[rgb]{0,0,0}Power}%
}}}}
\put(4583,-14468){\makebox(0,0)[b]{\smash{{\SetFigFont{7}{8.4}{\sfdefault}{\mddefault}{\updefault}{\color[rgb]{0,0,0}scaling}%
}}}}
{\color[rgb]{0,0,0}\put(3308,-21082){\framebox(2551,1890){}}
}%
\put(4583,-20042){\makebox(0,0)[b]{\smash{{\SetFigFont{7}{8.4}{\sfdefault}{\mddefault}{\updefault}{\color[rgb]{0,0,0}Power}%
}}}}
\put(4583,-20562){\makebox(0,0)[b]{\smash{{\SetFigFont{7}{8.4}{\sfdefault}{\mddefault}{\updefault}{\color[rgb]{0,0,0}scaling}%
}}}}
{\color[rgb]{0,0,0}\put(16583,-14987){\framebox(2551,1889){}}
}%
\put(17859,-13948){\makebox(0,0)[b]{\smash{{\SetFigFont{7}{8.4}{\sfdefault}{\mddefault}{\updefault}{\color[rgb]{0,0,0}MMSE}%
}}}}
\put(17859,-14468){\makebox(0,0)[b]{\smash{{\SetFigFont{7}{8.4}{\sfdefault}{\mddefault}{\updefault}{\color[rgb]{0,0,0}estimation}%
}}}}
{\color[rgb]{0,0,0}\put(16583,-21129){\framebox(2551,1890){}}
}%
\put(17859,-20090){\makebox(0,0)[b]{\smash{{\SetFigFont{7}{8.4}{\sfdefault}{\mddefault}{\updefault}{\color[rgb]{0,0,0}MMSE}%
}}}}
\put(17859,-20609){\makebox(0,0)[b]{\smash{{\SetFigFont{7}{8.4}{\sfdefault}{\mddefault}{\updefault}{\color[rgb]{0,0,0}estimation}%
}}}}
{\color[rgb]{0,0,0}\put(426,-14042){\vector( 1, 0){2882}}
}%
{\color[rgb]{0,0,0}\put(5859,-14042){\vector( 1, 0){5338}}
}%
{\color[rgb]{0,0,0}\put(8957,-20184){\vector( 1, 0){2193}}
}%
{\color[rgb]{0,0,0}\put(11658,-14049){\vector( 1, 0){4925}}
}%
{\color[rgb]{0,0,0}\put(11434,-15649){\vector( 0, 1){1375}}
}%
{\color[rgb]{0,0,0}\put(11611,-20190){\vector( 1, 0){4972}}
}%
{\color[rgb]{0,0,0}\put(5859,-20184){\vector( 1, 0){2598}}
}%
{\color[rgb]{0,0,0}\put(7087,-17586){\line( 1, 0){1654}}
\put(8741,-17586){\vector( 0,-1){2362}}
}%
{\color[rgb]{0,0,0}\put(4044,-17586){\vector( 1, 0){1201}}
}%
{\color[rgb]{0,0,0}\put(1182,-14042){\line( 0,-1){3544}}
\put(1182,-17586){\vector( 1, 0){992}}
}%
{\color[rgb]{0,0,0}\put(2174,-18531){\framebox(1842,1890){}}
}%
{\color[rgb]{0,0,0}\put(473,-20184){\vector( 1, 0){2835}}
}%
{\color[rgb]{0,0,0}\put(14221,-20184){\line( 0, 1){2598}}
\put(14221,-17586){\vector( 1, 0){1228}}
}%
{\color[rgb]{0,0,0}\put(19134,-20184){\vector( 1, 0){4961}}
}%
{\color[rgb]{0,0,0}\put(19182,-18531){\framebox(1842,1890){}}
}%
{\color[rgb]{0,0,0}\put(17272,-17586){\vector( 1, 0){1910}}
}%
{\color[rgb]{0,0,0}\put(21024,-17586){\vector( 1, 0){3071}}
}%
{\color[rgb]{0,0,0}\put(19134,-14042){\line( 1, 0){3307}}
\put(22441,-14042){\line( 0,-1){1418}}
\put(22441,-15460){\line(-1, 0){2362}}
\put(20079,-15460){\vector( 0,-1){1181}}
}%
{\color[rgb]{0,0,0}\put(6142,-19475){\vector( 0, 1){944}}
\put(6142,-19475){\line( 1, 0){1418}}
\put(7560,-19475){\line( 0,-1){709}}
}%
{\color[rgb]{0,0,0}\put(15449,-18531){\framebox(1843,1890){}}
}%
{\color[rgb]{0,0,0}\put(5245,-18531){\framebox(1842,1890){}}
}%
\put(1010,-19989){\makebox(0,0)[b]{\smash{{\SetFigFont{7}{8.4}{\familydefault}{\mddefault}{\updefault}{\color[rgb]{0,0,0}$S_2$}%
}}}}
\put(1010,-13847){\makebox(0,0)[b]{\smash{{\SetFigFont{7}{8.4}{\familydefault}{\mddefault}{\updefault}{\color[rgb]{0,0,0}$S_1$}%
}}}}
\put(11575,-15979){\makebox(0,0)[b]{\smash{{\SetFigFont{7}{8.4}{\familydefault}{\mddefault}{\updefault}{\color[rgb]{0,0,0}$Z_{s_1}$}%
}}}}
\put(11528,-22121){\makebox(0,0)[b]{\smash{{\SetFigFont{7}{8.4}{\familydefault}{\mddefault}{\updefault}{\color[rgb]{0,0,0}$Z_{s_2}$}%
}}}}
\put(3071,-17680){\makebox(0,0)[b]{\smash{{\SetFigFont{7}{8.4}{\sfdefault}{\mddefault}{\updefault}{\color[rgb]{0,0,0}source}%
}}}}
\put(3071,-18200){\makebox(0,0)[b]{\smash{{\SetFigFont{7}{8.4}{\sfdefault}{\mddefault}{\updefault}{\color[rgb]{0,0,0}encoder}%
}}}}
\put(20079,-17680){\makebox(0,0)[b]{\smash{{\SetFigFont{7}{8.4}{\sfdefault}{\mddefault}{\updefault}{\color[rgb]{0,0,0}source}%
}}}}
\put(20079,-18200){\makebox(0,0)[b]{\smash{{\SetFigFont{7}{8.4}{\sfdefault}{\mddefault}{\updefault}{\color[rgb]{0,0,0}decoder}%
}}}}
\put(23386,-17444){\makebox(0,0)[b]{\smash{{\SetFigFont{7}{8.4}{\sfdefault}{\mddefault}{\updefault}{\color[rgb]{0,0,0}$\widehat{S}_1$}%
}}}}
\put(23292,-20042){\makebox(0,0)[b]{\smash{{\SetFigFont{7}{8.4}{\sfdefault}{\mddefault}{\updefault}{\color[rgb]{0,0,0}$\widehat{S}_2$}%
}}}}
\put(7796,-13806){\makebox(0,0)[b]{\smash{{\SetFigFont{7}{8.4}{\familydefault}{\mddefault}{\updefault}{\color[rgb]{0,0,0}$(P_1)$}%
}}}}
\put(7796,-17397){\makebox(0,0)[b]{\smash{{\SetFigFont{7}{8.4}{\familydefault}{\mddefault}{\updefault}{\color[rgb]{0,0,0}$(P_2^\prime)$}%
}}}}
\put(7938,-20798){\makebox(0,0)[b]{\smash{{\SetFigFont{7}{8.4}{\familydefault}{\mddefault}{\updefault}{\color[rgb]{0,0,0}$(P_2-P_2^\prime)$}%
}}}}
\put(3071,-17255){\makebox(0,0)[b]{\smash{{\SetFigFont{7}{8.4}{\sfdefault}{\mddefault}{\updefault}{\color[rgb]{0,0,0}WZ}%
}}}}
\put(20079,-17255){\makebox(0,0)[b]{\smash{{\SetFigFont{7}{8.4}{\sfdefault}{\mddefault}{\updefault}{\color[rgb]{0,0,0}WZ}%
}}}}
\put(16347,-18011){\makebox(0,0)[b]{\smash{{\SetFigFont{7}{8.4}{\sfdefault}{\mddefault}{\updefault}{\color[rgb]{0,0,0}decoder}%
}}}}
\put(16347,-17491){\makebox(0,0)[b]{\smash{{\SetFigFont{7}{8.4}{\sfdefault}{\mddefault}{\updefault}{\color[rgb]{0,0,0}dirty-paper}%
}}}}
\put(6142,-18011){\makebox(0,0)[b]{\smash{{\SetFigFont{7}{8.4}{\sfdefault}{\mddefault}{\updefault}{\color[rgb]{0,0,0}encoder}%
}}}}
\put(6142,-17491){\makebox(0,0)[b]{\smash{{\SetFigFont{7}{8.4}{\sfdefault}{\mddefault}{\updefault}{\color[rgb]{0,0,0}dirty-paper}%
}}}}
\end{picture}%
}\\
(c)
\end{center}
\caption{Strong-user-optimal case (also
see~\cite[Fig.~2]{prabhakaranprupgrade08}): (a) separation scheme showing
successive refinement (SR) and superposition coding, (b) separation scheme
with Wyner-Ziv (W-Z) code and dirty-paper coding (DPC), (c) the hybrid
digital-analog scheme.}
\label{hybrid.fig:strong-user optimal}
\end{figure}
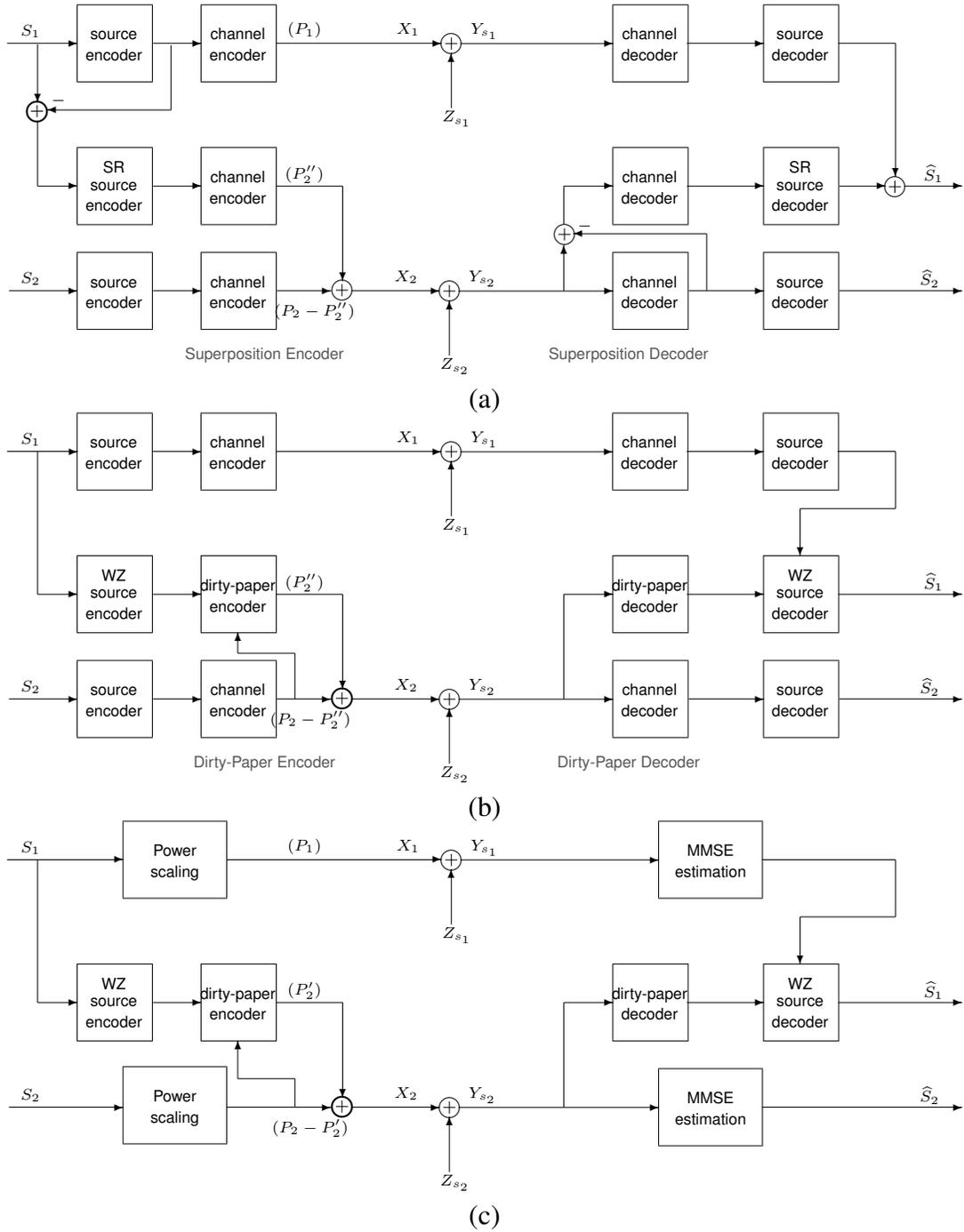

\begin{figure}[htb]
\begin{center}
\scalebox{0.8}{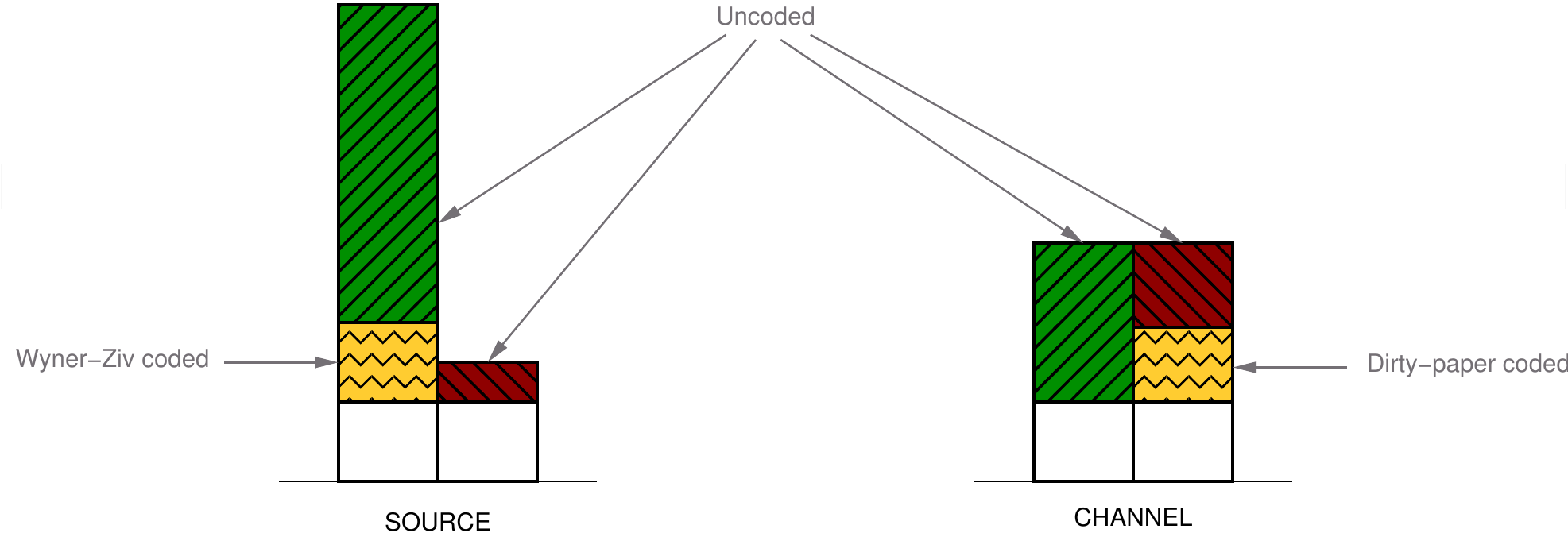}\\
\end{center}
\caption{
Strong-user-optimal case: The schematic diagram showing the allocation for
a $K=M=2$ example. The optimal coded separation scheme for the strong user
may be thought of as sending the coarse-layer bits from a successive
refinement source coding of the first source component $S_1$ alone over the
first sub-channel, and source coded bits of the second source component
$S_2$ and the refinement bits of $S_1$ over the second sub-channel.
However, none of these bits are decodable by the weak user. We may provide
useful information without compromising the strong user's performance by
(i) sending $S_1$ uncoded (scaled) over the first sub-channel and (ii)
sending $S_2$ uncoded (scaled) over the second sub-channel and bits
carrying refinement information on $S_1$ using Gel'fand and Pinsker's
(dirty-paper) channel coding where the transmission of $S_2$ acts as
Gaussian side-information at the transmitter. Dirty-paper coding ensures
that the transmission of $S_2$ does not affect the rate of transmission of
the refinement bits. The bits carrying the refinement information are
produced using Wyner-Ziv coding where the noisy observation of $S_1$ over
the first sub-channel acts as side-information at the decoder. While the
weak user will be unable to decode the refinement information, it benefits
from the two uncoded transmissions.}
\label{fig:suhistogram}
\end{figure}

It is again helpful to consider the special case of $K=M=2$ and $\sigma_1^2
> \sigma_2^2$ (see Fig.~\ref{fig:suhistogram}) before the general case.
Note that this example was also used as a starting point for presenting the
scheme in~\cite[Section~III]{prabhakaranprupgrade08}. We summarize the
discussion below for completeness. With $P_1=P_2=P$, we have
\[
\frac{1}{2}\log\left(\frac{\sigma_1^2}{D_1}\frac{\sigma_2^2}{D_2}\right) =
\frac{1}{2}\log\left(\frac{P_1+{N_s}}{{N_s}}\frac{P_2+{N_s}}{{N_s}}\right).\]
We source code $S_1$ using successive refinement (Fig.~\ref{hybrid.fig:strong-user optimal}(a)) such that now the bitrate
of the coarse description at distortion $D_1^{\prime\prime}$ is equal to
the rate at which the first sub-channel operates.
\[ \frac{1}{2}\log\left(\frac{\sigma_1^2}{D_1^{\prime\prime}}\right) =
\frac{1}{2}\log\left(1+\frac{P_1}{{N_s}}\right).\]
Note that this is different from the previous subsection where we set the
bitrate of the refinement bit stream equal to the rate of the first
sub-channel. Thus
\[ 
\frac{1}{2}\log\left(\frac{D_1^{\prime\prime}}{D_1}\frac{\sigma_2^2}{D_2}\right)
   = \frac{1}{2}\log\left(1+\frac{P_2}{{N_s}}\right).\]

Instead of sending the coarse description over the first sub-channel and the
refinement bit stream on the second, without losing optimality for the
strong user, we may send $S_1$ uncoded (scaled by $\sqrt{P_1/\sigma_1^2}$)
over the first sub-channel and on the second sub-channel a Wyner-Ziv bit
stream (of rate equal to that of the refinement bit stream) which assumes
that the corrupted version of $S_1$ from the first sub-channel will be
available as side information at the decoder. The optimality of this scheme
follows from the no rate-loss property of jointly Gaussian sources under
Wyner-Ziv coding~\cite{wynerzrf76}. 

Thus the weak user can now form an estimate of $S_1$ from its output $Y_1$
of the first sub-channel. It is also possible to provide the weak user with
an estimate of the second component without losing optimality for the
strong user. Let us define $P_2^{\prime\prime}$ such that the Wyner-Ziv bit
stream can be sent using this power, superposition coded with the bit
stream for $S_2$ which uses the rest of the power $P_2-P_2^{\prime\prime}$.
The decoding order is first the bit stream for $S_2$, followed by the
Wyner-Ziv bit stream. Thus
\[ \frac{1}{2}\log\left(\frac{D_1^{\prime\prime}}{D_1}\right)
 =\frac{1}{2}\log\left(1+\frac{P_2^{\prime\prime}}{{N_s}}\right),\]
and
\[ \frac{1}{2}\log\left(\frac{\sigma_2^2}{D_2}\right)
 =\frac{1}{2}\log\left(1+\frac{P_2-P_2^{\prime\prime}}{P_2^{\prime\prime}+{N_s}}\right).\]
We can use dirty-paper coding of Gel'fand and Pinsker~\cite{gelfandpcc80}
and Costa~\cite{costawd83} instead of superposition coding to achieve the
same rates. Here the channel codeword for the bit stream of $S_2$ is
treated as non-causal side information available at the encoder when
channel encoding the Wyner-Ziv bit stream. Fig.~\ref{hybrid.fig:strong-user
optimal}(b) shows this setup which uses a combination of dirty-paper coding
and Wyner-Ziv coding. Note that the decoder does not need to decode the bit
stream for $S_2$ in order to decode the Wyner-Ziv bit stream.  This allows
us to send $S_2$ uncoded (scaled by
$\sqrt{(P_2-P_2^{\prime\prime})/\sigma_2^2}$) without impacting the
optimality for the strong user (Fig.~\ref{hybrid.fig:strong-user
optimal}(c)). The weak user can now form an estimate of $S_2$ from $Y_2$.

Again, we can easily extend the above intuition to $K=M>2$. Let $L$ be the
number of source components and sub-channels such that
$(1/2)\log(\sigma_k^2/D_k) > (1/2)\log(1+P_k/N_s)$. Since $\sigma_k^2$ are
monotonically decreasing, these will be the first $L$ components. For these
components, we define $D_k^{\prime\prime}$ such that
$(1/2)\log(\sigma_k^2/D_k^{\prime\prime}) = (1/2)\log(1+P_k/N_s)$. For the
rest of the sub-channels ($m>L$) we define $P_k^{\prime\prime}$ by
$(1/2)\log(\sigma_m^2/D_m) =
(1/2)\log(1+(P_m-P_m^{\prime\prime})/(P_m^{\prime\prime}+N_s))$. The first
$L$ source components are sent uncoded scaled by $\sqrt{P_k/\sigma_k^2}$ on
their corresponding sub-channels, and rest of the source components are sent
uncoded scaled by $\sqrt{(P_k-P_k^{\prime\prime})/\sigma_k^2}$ on their
corresponding sub-channels. The first $L$ source components are Wyner-Ziv
source coded at rates of $(1/2)\log(D_k^{\prime\prime}/D_k)$ assuming the
availability at the decoder (strong user) of the noise corrupted versions
sent over the corresponding sub-channels. These source coded bits are sent
using dirty-paper coding over the rest of the sub-channels $m>L$. The
resulting distortion for the weak user is
\[ D_w=\frac{1}{K}\left(\sum_{k=1}^L \frac{\sigma_k^2}{1+\frac{P_k}{N_w}} +
\sum_{k=L+1}^K \frac{\sigma_k^2}{1+\frac{P_k-P_k^{\prime\prime}}{N_w}}\right)\]
which is strictly less than $\frac{1}{K}\sum_{k=1}^L \sigma_k^2$, the
distortion for the weak user in the separation approach under which no
information is decodable by this user. 

This scheme also directly extends to the bandwidth expansion ($K<M$)
and bandwidth contraction ($K>M$) scenarios. To summarize, we can state the
following
\begin{thm}
The distortion pair $(D_s^\ast,D_w)$ is achievable for the source-channel
coding problem in section~\ref{hybrid.sec:problemstatement}, where the
point-to-point optimal distortion $D_s^\ast$ for the strong-user and $D_w$
are as follows.
\begin{equation}
D_s^\ast = \frac{1}{K}\sum_{k=1}^K D_k, \mbox{   where }
   D_k=\left\{ \begin{array}{ll} 
                 \mu, &\mbox{if } \mu < \sigma_k^2,\\
                 \sigma_k^2, &\mbox{if } \mu \ge \sigma_k^2,
               \end{array} \right.
\end{equation}
where $\mu$ is chosen such that the total rate $(1/2)\sum_{k=1}^K
\log(\sigma_k^2/D_k)$ equals the capacity $C_s$ of the strong user's channel.
$C_s = \sum_{m=1}^M\log(1+P_m/{N_s})$ where $P_1=P_2=\ldots=P_M=P$.

Let
\[L=\min\left\{\left|\left\{k:\frac{\sigma_k^2}{D_k}\ge 1 +
\frac{P_k}{N_s}\right\}\right|,M\right\},\]
\[K^\prime=\min\left\{\left|\left\{k:\mu\le\sigma_k^2\right\}\right|,K\right\},\]
and the $P_k^{\prime\prime}$'s be defined by
\[\frac{\sigma_m^2}{D_m} =
1+\frac{P_m-P_m^{\prime\prime}}{P_m^{\prime\prime}+N_s}.\]
Then,
\[ D_w=\frac{1}{K}\left(\sum_{k=1}^L \frac{\sigma_k^2}{1+\frac{P_k}{N_w}} +
\sum_{k=L+1}^{K^\prime}
\frac{\sigma_k^2}{1+\frac{P_k-P_k^{\prime\prime}}{N_w}} +
\sum_{k=K^\prime}^K \sigma_k^2.\right)\]

\end{thm}

\begin{figure}[htb]
\begin{center}
\scalebox{0.8}{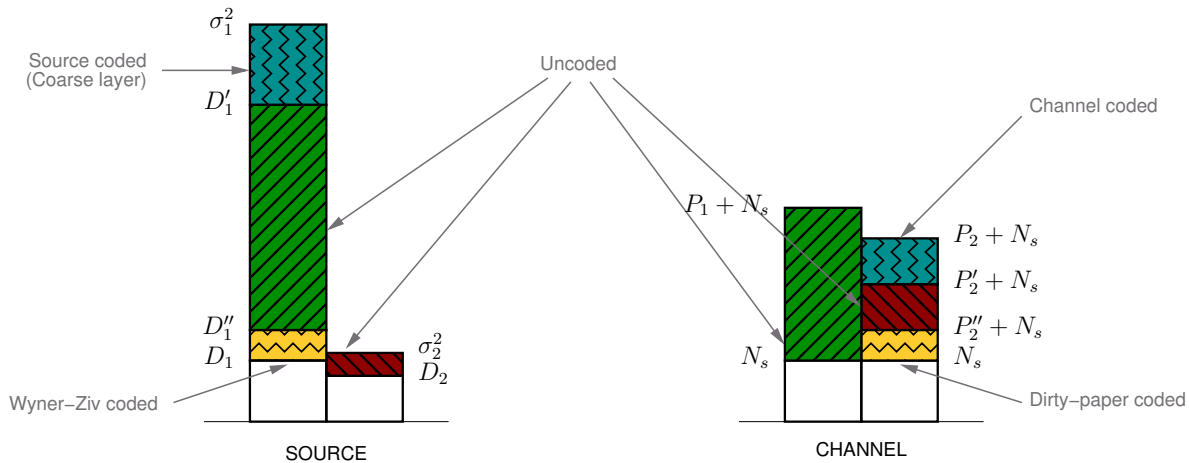}\\
\end{center}
\caption{
An achievable trade-off: The schematic diagram shows an allocation for a
$K=M=2$ example. The first source component $S_1$ is sent in three
different ways: (i) a coarse layer source codeword which will be decoded by
both users, (ii) an uncoded version of the quantization error which the
weak user will estimate from its noisy observation, and (iii) Wyner-Ziv
bits on $S_1$ which only the strong user decodes. In decoding the Wyner-Ziv
codeword, the strong user uses as side-information a linear estimate of
$S_1$ using (i) and its noisy observation of (ii). The quantization error
(ii) is sent uncoded (scaled) over the first sub-channel using all the
power allocated to this sub-channel. Over the second sub-channel, the
coarse layer bits from (i) above are sent using a Gaussian channel code
using part of the power allocated to this sub-channel. This is meant to be
decoded by both users treating the rest of the signals sent over this
sub-channel as noise. Using part of the leftover power, the second source
is sent uncoded (scaled). Since the codeword carrying bits from (i) is
assumed to be successfully decoded by both users, they may estimate the
second source component assuming only the rest of the power used in this
sub-channel and the channel noise as the disturbance affecting this
transmission. The leftover power in this sub-channel is used to send the
Wyner-Ziv bits from (iii). This is done using Gel'fand-Pinsker's
(dirty-paper) coding treating the scaled version of the second source
component sent over this sub-channel as side-information (at the
transmitter).}
\label{fig:histogram}
\end{figure}

\section{An achievable trade-off} \label{hybrid.sec:general}
We may also trade-off the quality of reproductions at the two users without
being optimal for either. Clearly, time sharing between the two achievable
extreme points is a possibility. It is often possible to do better. A natural
strategy suggested by the above discussion is to combine the schemes for the
weak- and strong-user-optimal cases.

Again, we will first consider a $K=M=2$ example with $\sigma_1^2 >
\sigma_2^2$. Let us suppose that the power allocation to the two
sub-channels are $P_1$ and $P_2$ such that $P_1+P_2=2P$. Consider
Fig.~\ref{fig:histogram}. The first source component $S_1$ is sent in three
different ways: 
\begin{enumerate}
\item[(i)] 
A (coarse layer) source codeword is formed which is meant to be decoded by
both users. The source coded bits are sent over the second sub-channel
using a Gaussian channel code utilizing part of the power $P_2$ allocated
to this sub-channel. This is meant to be decoded by both users treating the
rest of the signals sent over this sub-channel as noise. Let us denote the
quantized version of $S_1^n$ by $S_1'^n$ and MSE by $D_1'$. The rate of this
coarse layer source codebook is then
\begin{align}
\frac{1}{2}\log\left(\frac{\sigma_1^2}{D_1'}\right).\label{eq:R1coarse}
\end{align}

\item[(ii)] An uncoded (scaled) version of the quantization error resulting
from the source coding in (i) is transmitted. This transmission occurs over
the first sub-channel using all the power $P_1$ allocated to this
sub-channel, {\em i.e.}, 
\[X_1^n=\sqrt{\frac{P_1}{D_1'}}\left(S_1^n-S_1'^n\right).\]
The weak user estimates $S_1^n$ from the codeword $S_1'^n$ in (i) and its
noisy observation of the quantization error over the first sub-channel 
using a linear estimator
\[ \widehat{S}_{w_1}^n=\frac{P_1}{P_1+N_w}
    \left(\sqrt{\frac{D_1'}{P_1}}Y_1^n\right)+S_1'^n.
\]
The resulting MSE is
\begin{align}
\frac{D_1'}{1+\frac{P_1}{N_w}}.\label{eq:D1'}
\end{align}
\item[(iii)] Wyner-Ziv bits on $S_1^n$ are sent intended for the strong
user alone. In decoding the Wyner-Ziv codeword, the strong user uses as
side-information a linear estimate of $S_1^n$ it forms using (i) and its
noisy observation of (ii) in a manner similar to the weak user's estimate
of $S_1^n$ above. Let us denote the strong user's estimate of $S_1^n$ by
$S_1''^n$ and the MSE of this estimate by $D_1''$. We have
\begin{align}
 S_1''^n &=\frac{P_1}{P_1+N_s}
           \left(\sqrt{\frac{D_1'}{P_1}}Y_1^n\right)+S_1'^n\notag\\
         &=\frac{P_1}{P_1+N_s}S_1^n +
          \left(1-\frac{P_1}{P_1+N_s}\right)S_1^n +
          \frac{P_1}{P_1+N_s}\sqrt{\frac{D_1'}{P_1}}Z_{s_1}^n,\text{ and}
\label{eq:S1''}\\
D_1''&=\frac{D_1'}{1+\frac{P_1}{N_s}}.\label{eq:D1''}
\end{align}
Then, by Wyner-Ziv's theorem, the bit rate needed to achieve a distortion
of $D_1$ on the source $S_1^n$ at the strong user with $S_1''^n$ acting as
side-information is\footnote{However, note that by \eqref{eq:S1''}, our
side-information $S_1''^n$ is not $S_1^n$ corrupted by a memoryless
Gaussian disturbance as a classical statement of Wyner-Ziv's theorem would
require.  All we are guaranteed is that the side-information has a MSE of
$D_1''$ with respect to the source. A simple extension of the achievability
proof can handle this situation -- we may invoke the achievability part
with $S_1^n$ and $S_1''^n$ acting as the (vector) symbols; see, for
instance,~\cite[Appendix~IV]{reznicfzbe06}.\label{foot:WZ}}
\begin{align}
\frac{1}{2}\log\frac{D_1''}{D_1}.\label{eq:RWynerZiv}
\end{align}
The transmission of these Wyner-Ziv bits occurs over the second sub-channel
as described below.
\end{enumerate}
The transmission over the second sub-channel is meant to deliver the coarse
layer source codeword bits about $S_1$ and an estimate of $S_2$ to the weak
user, and in addition to these, the Wyner-Ziv bits about $S_1$ as well to
the strong user. This is accomplished as follows:
\begin{enumerate}

\item[(a)] The source codeword bits from (i) are transmitted using a
Gaussian channel code utilizing power $P_2-P_2'$ (a part of the total power
$P_2$ allocated to this sub-channel). This is decoded by both users
treating the rest of the signals sent over this sub-channel as noise. The
decoding at the weak user presents the bottleneck to the rate at which bits
may be delivered. Hence, to meet the rate required by~\eqref{eq:R1coarse},
we must have $D_1'$, $P_2$, and $P_2'$ satisfy
\begin{align}
\frac{1}{2}\log\left(\frac{\sigma_1^2}{D_1^\prime}\right) = 
  \frac{1}{2}\log\left(1+\frac{P_2-P_2^\prime}{P_2^\prime+N_w}\right).
\label{eq:miniconds_start}
\end{align}

\item[(b)] The second source component $S_2^n$ is sent uncoded (scaled)
using power $P_2'-P_2''$ (a part of the power $P_2'$ leftover after (a)),
{\em i.e.}, we send $\sqrt{(P_2'-P_2'')/\sigma_2^2}S_2^n$. Since the
codeword from (a) is assumed to be successfully decoded by both users, they
may strip it off their received signals $Y_{w_2}^n$ and $Y_{s_2}^n$ and
estimate the second source component assuming only the rest of the power
used in this sub-channel and the channel noise as the disturbance affecting
this transmission. Both users employ linear estimators.  Let us denote the
signals after the codeword from (a) has been stripped off by $Y_{w_2}'^n$
and $Y_{s_2}''^n$ at the weak user and the strong user, respectively. Then,
the estimates of $S_2$ are
\begin{align*}
\widehat{S}_{j_2}^n=\frac{P_2'-P_2''}{P_2'+N_j}
      \left(\sqrt{\frac{\sigma_2^2}{P_2'-P_2''}}Y_{j_2}'^n\right),\quad
j\in\{s,w\}.
\end{align*}
The MSE $D_2$ on the second source component incurred by the strong user is
given by
\begin{align}
 \frac{1}{2}\log\left(\frac{\sigma_2^2}{D_2}\right)=
   \frac{1}{2}\log\left(1+\frac{P_2^{\prime}-P_2^{\prime\prime}}{P_2^{\prime\prime}+N_s}\right).\label{eq:D2}
\end{align}
Similarly, the weak user incurs an average distortion on the second source
component of
\[ \frac{\sigma_2^2}{1+\frac{P_2^{\prime}-P_2^{\prime\prime}}{P_2^{\prime\prime}+N_w}}.\]

\item[(c)] Finally, the Wyner-Ziv bits from (iii) are transmitted intended
for the strong user with the leftover power of $P_2''$. Let us recall that
we have assumed that the codeword from (a) is successfully decoded by the
strong user and stripped off its received signal. The only other
disturbances affecting the transmission of the Wyner-Ziv bits are the
memoryless Gaussian noise in the channel and the scaled transmission of the
memoryless Gaussian $S_2^n$ in (b), both of which are independent of each
other and the Wyner-Ziv bits being sent. Also, the disturbance $S_2^n$ is
known to the transmitter non-causally. This is precisely the setting of
Gel'fand-Pinsker or dirty-paper coding by which a rate equal to the
capacity of channel in which only the memoryless Gaussian noise is present
can be achieved. This rate must be large enough to support the Wyner-Ziv
bits whose rate is~\eqref{eq:RWynerZiv}. Thus, $P_2''$, $D_1''$ and $D_1$
must satisfy
\begin{align} \frac{1}{2}\log\left(\frac{D_1^{\prime\prime}}{D_1}\right)=
   \frac{1}{2}\log\left(1+\frac{P_2^{\prime\prime}}{N_s}\right).
\label{eq:miniconds_end}
\end{align}
\end{enumerate}
To summarize, the decoders can achieve distortions of
\begin{align*}
D_s &= (D_1+D_2)/2,\text{ and}\\
D_w &= \frac{1}{2}\left( \frac{D_1^\prime}{1+\frac{P_1}{N_w}} +
            \frac{\sigma_2^2}{1+\frac{P_2^{\prime}-P_2^{\prime\prime}}{P_2^{\prime\prime}+N_w}}\right),
\end{align*}
for every choice of the non-negative power parameters $P_1$, and $P_2\geq
P_2'\geq P_2''$ which satisfy the sum power constraint $P_1+P_2=2P$, and
non-negative distortion parameters $D_1\leq D_1''\leq D_1'\leq
\sigma_1^2$, and $D_2\leq \sigma_2^2$ provided they satisfy the conditions
\eqref{eq:D1'}, \eqref{eq:D1''}, \eqref{eq:miniconds_start}, \eqref{eq:D2},
and \eqref{eq:miniconds_end}.

Generalizing the above, in general, we have the following achievable trade-off
\begin{thm} \label{hybrid.thm:general}

Let $L\in\{0,1,\ldots,\min(K,M)\}$ and $K^\prime\in \{L, L+1, \ldots,
\min(K,M)\}$. Also, let $P_1, P_2, \ldots$, $P_M$, $P_{L+1}', P_{L+2}',
\ldots$, $P_M'$, $P_{L+1}'', P_{L+2}'', \ldots, P_{K'}''$, $D_1, D_2,
\ldots, D_M$, $D_1^\prime, D_2^\prime, \ldots, D_L$, $D_{K'+1}', D_{K'+2}',
\ldots, D_K$, $D_1^{\prime\prime}$, $D_2^{\prime\prime}, \ldots$,
$D_L^{\prime\prime}$ be non-negative such that the following conditions are
satisfied

\begin{align}
\sum_{m=1}^M P_m &\leq MP,\label{hybrid.eq:condstart}\\
P_m'&\leq P_m,\;m=L+1,L+2,\ldots,M,\notag\\
P_m''&\leq P_m',\;m=L+1,L+2,\ldots,K',\notag\\
D_k\leq D_k^{\prime\prime}\leq D_k^\prime &\leq
\sigma_k^2,\;k=1,2,\ldots,L,\notag\\
D_k &\leq \sigma_k^2,\;k=L+1,L+2,\ldots,K',\notag\\
D_k\leq D_k^\prime &\leq \sigma_k^2,\;k=K'+1,K'+2,\ldots,K.
\end{align}
The following $(D_s,D_w)$ is achievable
\begin{align*}
D_s&=\frac{1}{K}\left(\sum_{k=1}^K D_k\right), \\
D_w&=\frac{1}{K}\left(\sum_{k=1}^L \frac{D_k^\prime}{1+\frac{P_k}{N_w}} + 
\sum_{k=L+1}^{K^\prime}\frac{\sigma_k^2}{1+\frac{P_k'-P_k''}{P_k''+N_w}}
+ \sum_{k=K^\prime+1}^K D_k^{\prime}\right),
\end{align*}
if the following conditions are satisfied
\begin{align}
\frac{D_k'}{D_k''}&=1+\frac{P_k}{N_s},\quad k=1,2,\ldots, L,\notag\\
\frac{\sigma_k^2}{D_k}&={1+\frac{P_k'-P_k''}{P_k''+N_s}},\quad k=L+1, L+2,
\ldots, K',\notag\\
\sum_{k=1}^L \log\frac{\sigma_k^2}{D_k^\prime}  + \sum_{k=K^\prime+1}^K
\log \frac{\sigma_k^2}{D_k^\prime}
& \leq \sum_{m=L+1}^M
\log\left( 1 + \frac{P_m-P_m'}{P_m'+N_w}\right),\mbox{ and}
\label{hybrid.eq:weakuserdigitalcond}
\\
\sum_{k=1}^L 
\log\frac{D_k^{\prime\prime}}{D_k}  +
\sum_{k=K^\prime+1}^K  \log\frac{D_k^\prime}{D_k}
& \leq
\sum_{m=L+1}^{K^\prime} \log\left(1+\frac{P_m''}{N_s}\right)
+\sum_{m=K^\prime+1}^M \log\left(1+\frac{P_m'}{N_s}\right).
\label{hybrid.eq:stronguserdigitalcond}
\end{align}
\end{thm}
The proof is relegated to appendix~\ref{hybrid.app:general}.

\begin{figure}[htb]
\begin{center}
\scalebox{0.8}{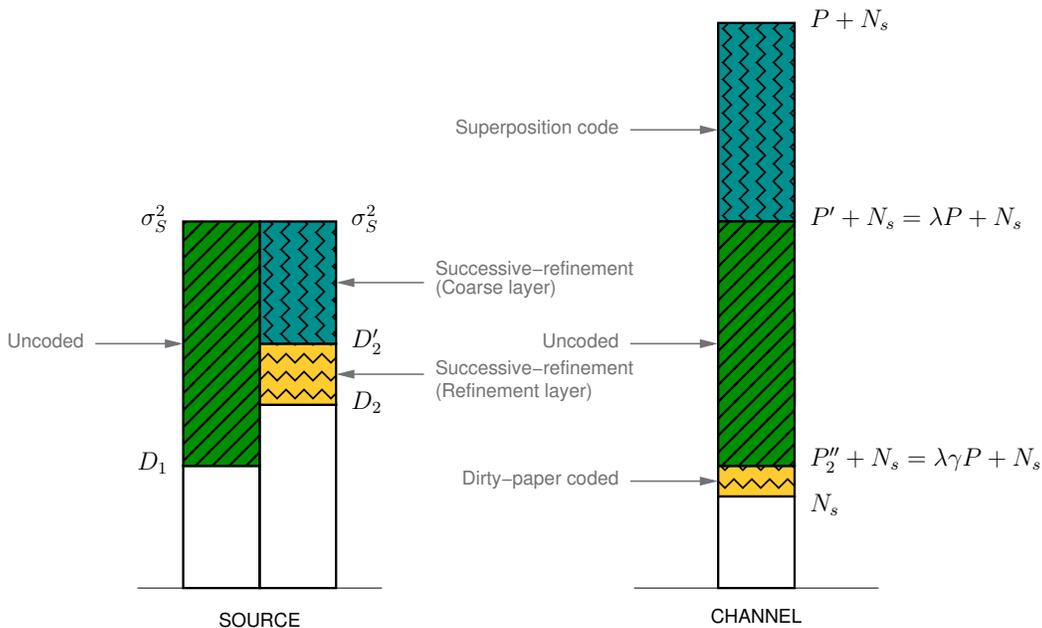}\\
\caption{
Memoryless source over memoryless channel with bandwidth contraction: 
Schematic diagram showing an allocation for bandwidth expansion factor
$\alpha=1/2$. Note that we need not use Wyner-Ziv coding and may use
successive refinement source coding as shown.
}
\label{fig:contracthistogram}
\end{center}
\end{figure}
\begin{figure}[htb]
\begin{center}
\scalebox{0.8}{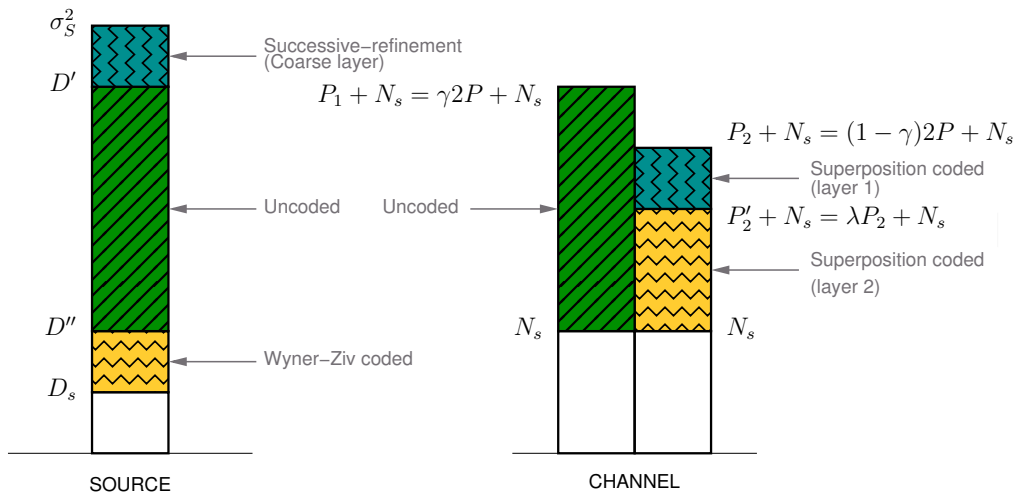}\\
\caption{
Memoryless source over memoryless channel with bandwidth expansion: 
Schematic diagram showing an allocation for bandwidth expansion factor
$\alpha=2$. Note that we need not use Gel'fand-Pinsker's (dirty-paper)
coding and may use superposition channel coding as shown.
}
\label{fig:expandhistogram}
\end{center}
\end{figure}

In general, the above optimization problem appears to be computationally
challenging for large values of $K$ and $M$. However, simplification is
possible for the important special case of memoryless sources and channels
with bandwidth expansion and bandwidth contraction. This is taken up in the
next section.

\section{Specialization to memoryless sources and channels with bandwidth
mismatch} \label{hybrid.sec:bandwidthmismatch}

A special case of the problem is when the source is also memoryless, but has a
bandwidth different from the channel. If we define the degree of mismatch
by $\alpha=M/K$, we have 
\begin{thm} \label{hybrid.thm:bandwidthmismatch}
For the special case of the problem in section~\ref{hybrid.sec:problemstatement}
with $\sigma_1^2=\sigma_2^2=\ldots=\sigma_M^2=\sigma_S^2$, the following $(D_s,D_w)$
trade-off is achievable:
\begin{itemize}
\item For $\alpha<1$ (bandwidth contraction)
\[ \left\{ \left( D^{\text{BC}}_s(\lambda,\gamma),
D^{\text{BC}}_w(\lambda,\gamma)\right):
0\leq\lambda\leq1,0\leq\gamma\leq1\right\},\]
where
\begin{align*}
D^{\text{BC}}_w(\lambda,\gamma) &=
\frac{\alpha\sigma_S^2}{\frac{\lambda P+N_w}{\lambda\gamma P+N_w}}
+ \frac{(1-\alpha)\sigma_S^2}{\left(\frac{P+N_w}{\lambda P +
N_w}\right)^{\frac{\alpha}{1-\alpha}}},\mbox{ and}\\
D^{\text{BC}}_s(\lambda,\gamma) &=
\frac{\alpha\sigma_S^2}{\frac{\lambda P+N_s}{\lambda\gamma P+N_s}}
+ \frac{(1-\alpha)\sigma_S^2}{\left(\frac{P+N_w}{\lambda P +
N_w}\frac{\lambda\gamma P+N_s}{N_s}\right)^{\frac{\alpha}{1-\alpha}}}.
\end{align*}
\item For $\alpha>1$ (bandwidth expansion)
\[ \left\{ \left( D^{\text{BE}}_s(\lambda,\gamma),
 D^{\text{BE}}_w(\lambda,\gamma) \right): 0\leq\lambda\leq1,\;
0\leq \gamma\leq 1\right\},\]
where
\begin{align*}
D^{\text{BE}}_w(\lambda) &=
\frac{\sigma_S^2}{\left(\frac{\frac{\alpha(1-\gamma)}{\alpha-1}P+N_w}
{\lambda\frac{\alpha(1-\gamma)}{\alpha-1}P + N_w}\right)^{\alpha-1}
\left(\frac{\alpha\gamma P+N_w}{N_w}\right)},\mbox{ and}\\
D^{\text{BE}}_s(\lambda) &=
\frac{\sigma_S^2}{\left(\frac{\frac{\alpha(1-\gamma)}{\alpha-1}P+N_w}
{\lambda\frac{\alpha(1-\gamma)}{\alpha-1}P + N_w}\right)^{\alpha-1}
\left(\frac{\alpha\gamma P+N_s}{N_s}\right)
\left(\frac{\lambda\frac{\alpha(1-\gamma)}{\alpha-1}P+N_s}{N_s}\right)^{\alpha-1}}.
\end{align*}
\end{itemize}
\end{thm}
We prove this as a special case of Theorem~\ref{hybrid.thm:general} in
appendix~\ref{hybrid.app:bandwidthmismatch}.

\begin{figure}[htb]
\begin{center}
\scalebox{0.7}{\includegraphics{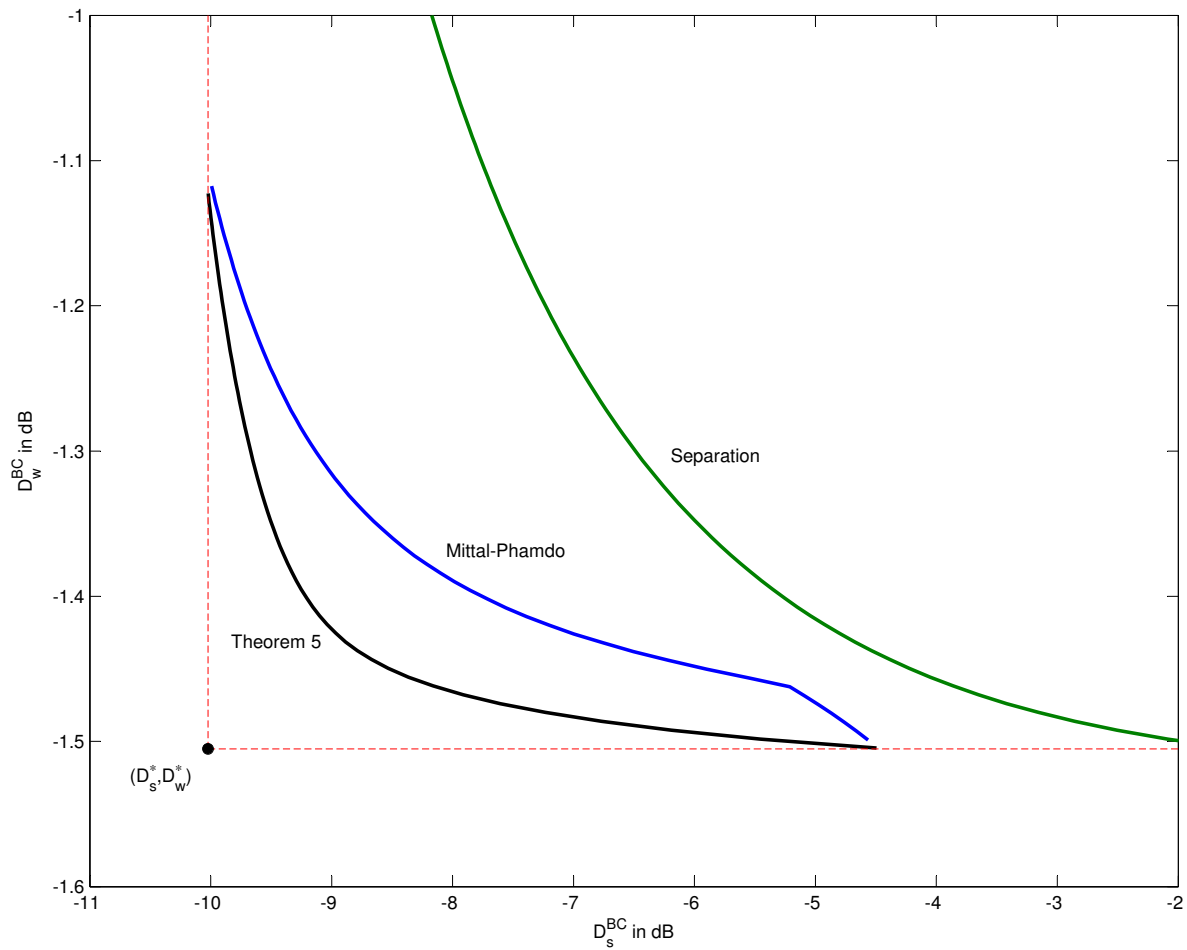}}\\
(a)
\end{center}
\caption{
Comparison of distortion trade-offs achieved by our scheme with that of
Mittal and Phamdo: (a) Bandwidth contraction.
$\sigma_S^2=1,\,P/N_s=20\text{dB},\,P/N_w=0\text{dB},\,\alpha=0.5$. The
best scheme suggested by Mittal and Phamdo was chosen for
comparison~\cite[Fig.  14]{mittalphd2002}. The dashed lines are drawn at
the weak and strong user optimal distortions and thus give the trivial
outer bound to the trade-off region. The strong-user-optimal points appear
to coincide, but there is a small gap which is not visible at the scale of
this plot. The weak user optimal points indeed coincide where both schemes
reduce to the same scheme. The performance of the best separation-based
scheme is also shown for comparison.
}
\label{hybrid.fig:whitesourcecomparison}
\end{figure}

For bandwidth contraction, we use only successive refinement source coding
and Gel'fand-Pinsker channel coding; see Fig.~\ref{fig:contracthistogram}.
In the bandwidth expansion case, only Wyner-Ziv coding and superposition
decoding is used (Fig.~\ref{fig:expandhistogram}). As pointed out earlier,
many researchers have investigated this special case.  For the bandwidth
expansion case, an almost identical scheme (using Wyner-Ziv coding and
superposition coding) was presented in~\cite{reznicfzbe06}.  The trade-off
expression above, but with $\gamma=1/\alpha$ (corresponding to a flat power
allocation) appears in~\cite[Theorem~2]{reznicfzbe06}. However, as
shown in Fig.~\ref{hybrid.fig:whitesourcecomparison}(b) and (c), the extra
flexibility from non-flat power allocations can lead to slight gains. The
bandwidth contraction case discussed above is new. The following remarks on
the extreme points of these trade-offs are in order.  
\begin{itemize}

\item At the weak-user-optimal points, our achievable schemes under bandwidth
contraction and bandwidth expansion reduce to the schemes proposed by
Mittal and Phamdo \cite{mittalphd2002}.

\item At the strong-user-optimal point under bandwidth expansion, our
scheme reduces to the systematic lossy source-channel codes of Shamai,
Verd\`{u}, and Zamir~\cite{shamaivzsl98}. 

\item 
At the strong-user-optimal points, the achievable scheme is strictly better
than the solution offered by Mittal and Phamdo in~\cite{mittalphd2002}. The
gap can be computed explicitly to be
\begin{align*}
\text{(bandwidth contraction)}&&&\frac{\alpha\sigma_S^2N_s}{N_w+P} \left(1
- \frac{1}{\left(1+\frac{P}{N_s}\right)^\alpha}\right),\quad\text{and}\\
\text{(bandwidth expansion)}&&&\frac{\sigma_S^2}{\left(1+\frac{P}{N_s}\right)^\alpha\left(1+\frac{N_w}{P}\right)}.
\end{align*}
As pointed out above, at the weak-user-optimal point, the schemes coincide.
For the boundary points in between the strong-user-optimal and
weak-user-optimal points, an explicit computation is cumbersome, but
numerical computation over a wide range of settings suggest that the
achievable scheme strictly out performs the schemes of Mittal and Phamdo.
Fig.~\ref{hybrid.fig:whitesourcecomparison} shows a comparison of the trade-offs
achieved by our scheme with those of Mittal and Phamdo~\cite{mittalphd2002}
for a few examples.
\end{itemize}

\addtocounter{figure}{-1}
\begin{figure}[htb]
\begin{center}
\scalebox{0.7}{\includegraphics{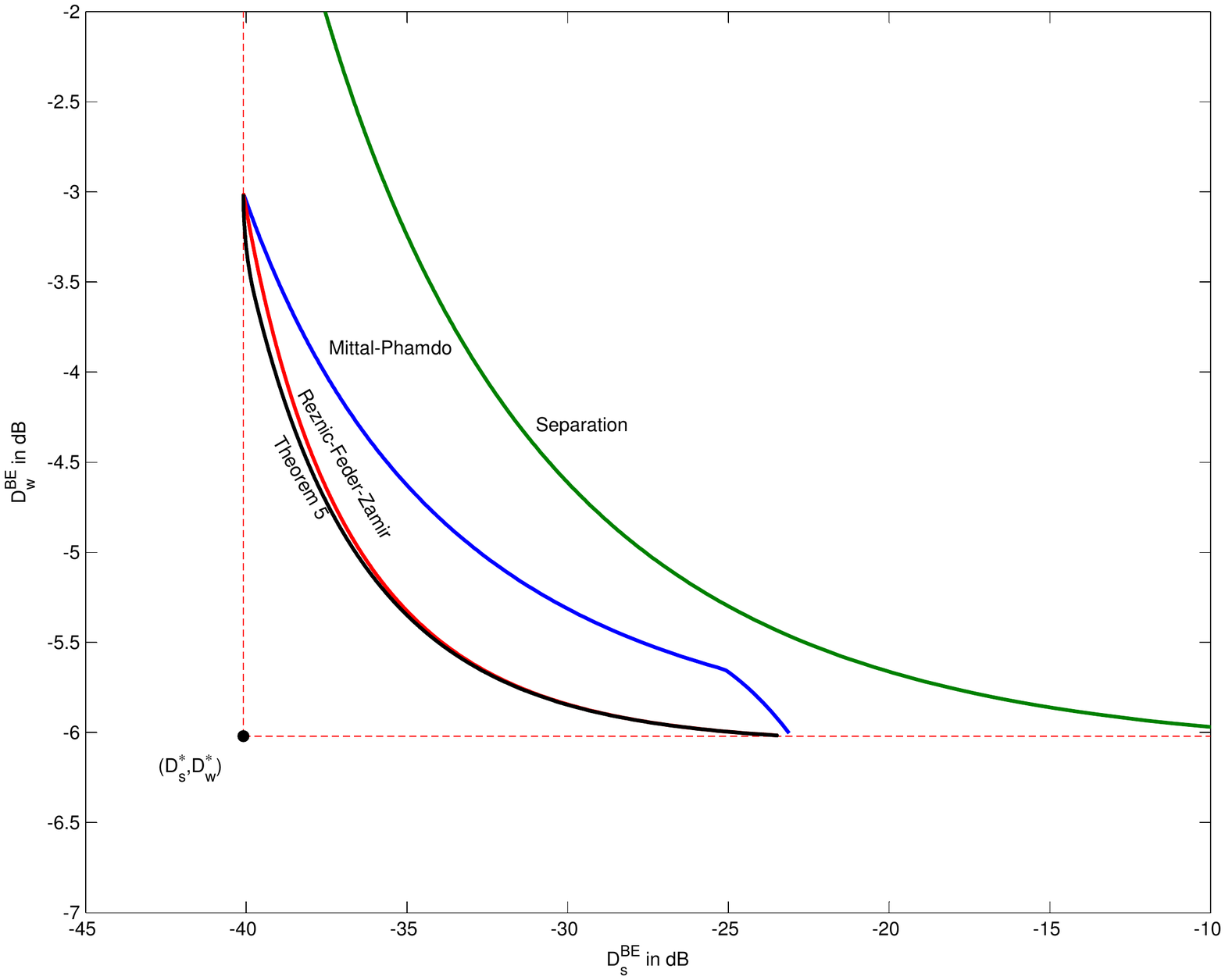}}\\
(b)
\end{center}
\caption{Comparison of distortion trade-offs achieved by our scheme with
that of Mittal and Phamdo: (b) Bandwidth expansion.
$\sigma_S^2=1,\,P/N_s=20\text{dB},\,P/N_w=0\text{dB},\,\alpha=2.0$. Again
the best scheme of Mittal and Phamdo for this setting was chosen for
comparison~\cite[Fig. 12]{mittalphd2002}. The dashed lines are drawn at the
weak and strong-user-optimal distortions and thus give the trivial outer
bound to the trade-off region. The strong-user-optimal points appear to
coincide, but there is a small gap which is not visible at the scale of
this plot. The weak-user-optimal points indeed coincide where the schemes
reduce to the same scheme. The slight improvement over~\cite{reznicfzbe06}
from allowing non-flat power allocation is visible. The performance of the
best separation-based scheme is shown for comparison.}
\end{figure}
\addtocounter{figure}{-1}
\begin{figure}[htb]
\begin{center}
\scalebox{0.7}{\includegraphics{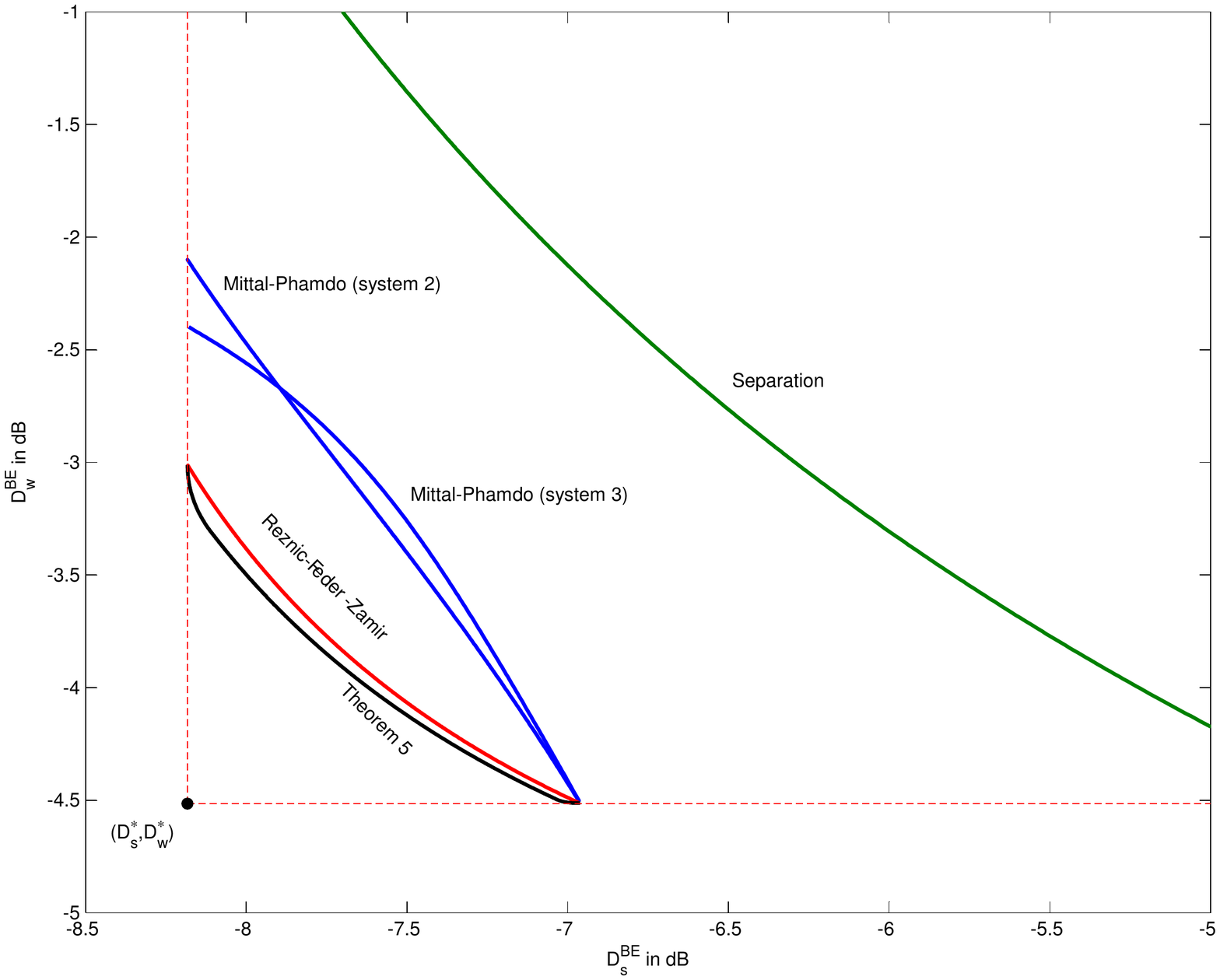}}\\
(c)
\end{center}
\caption{Comparison of distortion trade-offs achieved by our scheme with
those of Mittal and Phamdo: (b) Bandwidth expansion.
$\sigma_S^2=1,\,P/N_s=4\text{dB},\,P/N_w=0\text{dB},\,\alpha=1.5$. Two
schemes of Mittal and Phamdo (systems 2 and 3) together give the best
performance of all the new schemes proposed in~\cite[Fig.
13]{mittalphd2002}. The dashed lines are drawn at the weak and strong user
optimal distortions and thus give the trivial outer bound to the trade-off
region. The gap between the strong-user-optimal points is visible in this
plot. The weak-user-optimal points coincide. The slight improvement
over~\cite{reznicfzbe06} from allowing non-flat power allocation is again
visible. The performance of the best separation-based scheme is shown for
comparison.
}
\end{figure}

\noindent{\em Outer bounds:}
Only one non-trivial outer bound is available in the literature for this
problem. It is due to Reznic, Feder, and Zamir~\cite{reznicfzbe06} who
developed it for the case of memoryless source and channel under bandwidth
expansion. This bound, however, does not match the best available inner
bound described above. The same bounding technique can be used to derive
outer bounds for the parallel source-channel problem considered here.
However, it does not always lead to a non-trivial bound.  For instance, for
memoryless source and channel with bandwidth contraction, the technique
yields the trivial bound (resulting from considering the point-to-point
source-channel problems involving either the weak user or the strong users
alone).

\section{Conclusion} \label{hybrid.sec:conclusion}
We have presented a hybrid digital-analog scheme for the problem of sending a
parallel Gaussian source over a white Gaussian broadcast channel which
potentially has a bandwidth mismatch with the source. We used the concepts of
successive refinement and Wyner-Ziv source coding, and superposition and dirty
paper channel coding to show that without compromising the point-to-point
optimal performance of either the weak or strong user, we can strictly improve
the performance of the other user over what the conventional separation
approach offers.  We also showed how to achieve better trade-offs when neither
user is point-to-point optimal.  While we do not have a converse for our
scheme, the achievable points are the best available. When specialized to the
case of memoryless sources and channels with bandwidth mismatch, our scheme matches
or in some cases outperforms the best known schemes.

Only an achievable trade-off is available for two digital receivers with
different channels. We believe that the limitation is primarily due to the lack
of good outer bounds on the region of distortions that can be supported
simultaneously. Even in simple cases like when the source and channel are
memoryless, but with mismatching bandwidths, a tight result is not available.
In fact, the best available outer bound for the case where the source bandwidth
is larger than the channel bandwidth is the trivial outer bound which considers
the receivers separately. It is also not clear if codes with more structure can
be used to obtain better trade-offs. These could be subjects of further
investigation.

\section*{Acknowledgement}

The authors would like to thank the reviewers for their careful reading and
their suggestions which have helped improve the quality of this paper.

\begin{appendices}

\section{Proof of Proposition~\ref{prop:intro}}\label{app:gap}

We will use the notation introduced in
section~\ref{hybrid.sec:problemstatement} with $\sigma_1^2= \sigma_2^2 =
\ldots = \sigma_K^2 = \sigma^2$. The proposition is a consequence of the
following two facts: (i) the Gaussian source $\{S_k,\; k=1,\ldots,K\}$ is
successively refinable~\cite{equitzcsr91}, and (ii) there is an operating
point in the degraded message set rate region of the white Gaussian
broadcast channel in which each user's rate is within 1-bit of its
point-to-point capacity. To see the second fact, let us recall that the
boundary of the degraded message set rate region (for each sub-channel) of
the Gaussian broadcast channel is given by the set of rates
$\{(R_{\text{base}}(\beta),R_{\text{refine}}(\beta)),\; \beta\in[0,1]\}$,
where the rate of the message decoded by both users is  (in bits per
symbol)
\begin{align*}
R_{\text{base}}(\beta)&=\frac{1}{2}\log_2\left(1+\frac{(1-\beta)P}{\beta P
+N_w}\right),
\end{align*}
and the rate of the message decoded only by the strong user is
\begin{align*}
R_{\text{refine}}(\beta)&=\frac{1}{2}\log_2\left(1+\frac{\beta P}{N_s}\right).
\end{align*}
And, for a given $\beta\in[0,1]$, the overall rate delivered to the users is
\begin{align*}
R_w(\beta)&=R_{\text{base}}(\beta),\\
R_s(\beta)&=R_{\text{base}}(\beta)+R_{\text{refine}}(\beta).
\end{align*}
In order to show fact (ii), it is enough to show that there is a
$\bar{\beta}\in[0,1]$ such that 
\begin{align*}
R_w(\bar{\beta})+1&\geq R_w^\ast\defineqq R_w(0), \text{and}\\
R_s(\bar{\beta})+1&\geq R_s^\ast\defineqq R_s(1),
\end{align*}
where $R_w^\ast$ and $R_s^\ast$ are the point-to-point capacities
of the weak and strong users respectively. Simplifying the above two
conditions, we get
\[ \frac{N_w}{P}\geq\bar{\beta}\geq
\frac{1}{1/\left(\frac{N_w-N_s}{P}\right) +
2/\left(1+\frac{N_s}{P}\right)} - \frac{N_s}{P}.\]
Observing that the left-hand side is always larger than the right-hand
side and that the right-hand side is always less than 1, we can conclude that 
that such a $\bar{\beta}\in[0,1]$ always exists. By choosing this operating
point $(R_\text{base}(\bar{\beta}),R_\text{refine}(\bar{\beta}))$ for the
degraded message set broadcast channel code, and using the optimal layered
(successive refinement) source code, the distortion pair $(D_s,D_w)$ achieved
by the user-$j$, $j\in\{s,w\}$ satisfies
\begin{align*}
\frac{K}{2}\log_2\frac{\sigma^2}{D_j}&=MR_j(\bar{\beta})\\
&\geq M(R_j^\ast - 1)\\
&= \frac{K}{2}\log_2\frac{\sigma^2}{D_j^\ast} - M.
\end{align*}
Thus, since $M/K$ is the ratio of the channel bandwidth to the source
bandwidth, we have proved that
\[\frac{1}{2}\log_2\frac{D_j}{D_j^\ast}\leq\frac{\text{BW}_\text{channel}}{\text{BW}_\text{source}},\;j\in\{s,w\}.\]

\section{Proof of Theorem~\ref{hybrid.thm:bandwidthmismatch}}
\label{hybrid.app:bandwidthmismatch} 

The proof for the bandwidth contraction case ($\alpha=M/K<1$) follows from
the following choice of parameters in Theorem~\ref{hybrid.thm:general}: $L =
0,\; K^\prime=M$,
and the power allocation is
\begin{align*}
P_m&=P,\quad m=1,2,\ldots,M,\\
P_m'&=\lambda P,\quad m=1,2,\ldots,M,\\
P_m''&=\lambda\gamma P,\quad m=1,2,\ldots,M,
\end{align*}
where $\lambda$ and $\gamma$ are in $[0,1]$.
Also, we let
\begin{align*}
D_k&=D,\quad k=1,2,\ldots,M,\\
D_k&=\widetilde{D},\quad k=M+1,M+2,\ldots,K,\\
D_k^\prime&=\widetilde{D}^\prime,\quad k=M+1,M+2,\ldots,K,
\end{align*}
where $D, \widetilde{D}$, and $\widetilde{D}'$ are defined as
below to satisfy the conditions of Theorem~\ref{hybrid.thm:general}.
\begin{align*}
\frac{\sigma_S^2}{D}&=1+\frac{\lambda (1-\gamma) P}{\lambda\gamma P +
N_s},\\
(K-M)\log\frac{\sigma_S^2}{\widetilde{D}^\prime} &=
    M\log\frac{P+N_w}{\lambda P+N_w},\\
(K-M)\log\frac{\widetilde{D}^\prime}{\widetilde{D}} &=
    M\log\frac{\lambda\gamma P+N_s}{N_s}.
\end{align*}
Substituting these in the expression for the achievable $(D_s,D_w)$ gives
the result.

The choice of parameters for the bandwidth expansion case ($\alpha=M/K<1$)
is $L=K^\prime=K$, and
the power allocations are
\begin{align*}
P_m&=\frac{1}{K}(\gamma MP)=\alpha\gamma P,\quad m=1,2,\ldots,K,\\
P_m&=\frac{1}{M-K}((1-\gamma)MP)=\frac{\alpha(1-\gamma)}{\alpha-1}P, 
 \quad m=K+1,K+2,\ldots,M,
\end{align*}
where $\gamma\in[0,1]$. Clearly, we have $\sum_{m=1}^M P_m = MP$ as
required. Also, we set
\begin{align*}
P_m'&=(1-\lambda)P_m, \quad m=K+1,K+2,\ldots,M,
\end{align*}
where $\lambda\in[0,1]$, and $D_k'=D'$, $D_k''=D''$, $D_k=D$, 
$k=1,2,\ldots,K$, where $D'$, $D''$, and $D$ are chosen as follows to
satisfy the conditions in Theorem~\ref{hybrid.thm:general}.
\begin{align*}
K\log\frac{\sigma_S^2}{D'} &= (M-K)\log\frac
{\frac{\alpha(1-\gamma)}{\alpha-1}P+N_w}
{\lambda\frac{\alpha(1-\gamma)}{\alpha-1}P+N_w},\\
\frac{D'}{D''}&=1+\frac{\alpha\gamma P}{N_s},\\
K\log\frac{D''}{D} &= (M-K)\log\frac{\lambda
\frac{\alpha(1-\gamma)}{\alpha-1}P+N_s}{N_s}.
\end{align*}
These choices give the achievability result for bandwidth expansion.

\section{Proof of Theorem~\ref{hybrid.thm:general}} \label{hybrid.app:general}

The main ideas involved have already been described in
section~\ref{hybrid.sec:general}. We sketch the main steps of the proof
which use the following results: successive refinement source
coding~\cite{equitzcsr91}, source coding with side-information or Wyner-Ziv
(WZ) coding~\cite{wynerzrf76}, super-position broadcast channel
coding~\cite{coverbc72}, and channel coding with side-information or
Gel'fand-Pinsker (GP) coding~\cite{gelfandpcc80} (in particular, as applied
by Costa to the Gaussian case~\cite{costawd83}).

The $m$-th sub-channel is allocated a power of $P_m$ such that it satisfies
the power constraint by \eqref{hybrid.eq:condstart}. The coding will be
performed as usual on block-length $n$ sequences of sufficient length that
the source codes invoked below have distortions close to optimal and the
channel codes have low probabilities of errors. For clarity, we will
supress these small gaps as we did in the discussion of the $K=M=2$ example
in section~\ref{hybrid.sec:general}.

\paragraph{Source components 1 through $L$} These source components are
treated in a way similar to the first source component was in the $K=M=$ example
(Fig.~\ref{fig:histogram}) of section~\ref{hybrid.sec:general} in that they
are sent in three different ways. An $n$-length block of the $k$-th such
source component is source-coded (quantized) using an optimal source-code
at distortion $D_k^{\prime}$. This codeword will be made available to both
the strong and the weak receivers.  The rate required to do this is
\[ \sum_{k=1}^L \frac{1}{2}\log\frac{\sigma_k^2}{D_k^{\prime}}.\]
Let the decoder reconstruction of the $i$-th sample be
${S}_k^{\prime}(i)$.

The quantization error of the $k$-th source component is transmitted over
the $k$-th sub-channel using power $P_k$. In other words, the input to the
$k$-th sub-channel is
\[ X_k(i) = \sqrt{\frac{P_k}{D_k^{\prime}}}
(S_k(i)-{S}_k^{\prime}(i)),\;i=1,2,\ldots,n.\]

To produce its reconstruction, the weak-user adds to ${S}_k^\prime(i)$
the linear least-squared error estimate of $S_k(i)-{S}_k^\prime(i)$
from $Y_{w_k}(i)$
\[
\widehat{S}_{w_k}(i)=\frac{P_k}{P_k+N_w}
    \left(\sqrt{\frac{D_k'}{P_k}}Y_k(i)\right)+S_k'(i).
\]
This estimate is at an MSE estimation error of
${D_k^{\prime}}/{\left(1+\frac{P_k}{N_w}\right)}$. This gives the first term in the
expression for $D_w$ in the theorem. The strong user also performs the same
to get an intermediate reconstruction of the source component at MSE
distortion of $D_k''={D_k^{\prime}}/{\left(1+\frac{P_k}{N_s}\right)}$. This will act
as a side-information available at the decoder for a Wyner-Ziv source
coding of $S_k$ (again footnote~\ref{foot:WZ} applies). We would like to
enhance this to a distortion of $D_k$ in the expression for $D_s$ in the
theorem. Using an extension of Wyner and Ziv's result, the Wyner-Ziv
bitrate needed to be delivered to the strong user is 
\[ \sum_{k=1}^L
\frac{1}{2}\log\frac{D_k''}{D_k}.
\]

\paragraph{Source components $L+1$ through $K^\prime$} These source
components are similar to the second source component of the $K=M=$ example
(Fig.~\ref{fig:histogram}) of section~\ref{hybrid.sec:general} in that they
are sent uncoded, but the sub-channels over which they are sent may have
other bits sent using codewords.
The $k$-th
source component is sent uncoded on the $k$-th sub-channel using power
$P_k'-P_k''$. The rest of the power spent on this sub-channel
is utilized to send the coded parts as will be discussed below. However, we
need to note the fact that both decoders will be able to decode a coded part
sent using power $P_k-P_k'$ on this sub-channel and subtract it off
before estimating $S_k$. The rest of the coded part which is sent at power
$P_k''$ will act as interference. Hence the MSE distortion for the $k$-th
source component will be
\[
{D_j}_k=\frac{\sigma_k^2}{1+\frac{P_k'-P_k''}{P_k''+N_j}},\quad
j\in\{s,w\}.\]
This gives the corresponding terms in the expressions for $D_s$ and
$D_w$ in the theorem.

\paragraph{Source components $K^\prime+1$ through $K$} These source
components will have no uncoded component unlike the above two cases. Thus
they are source coded using an optimal successive refinement
code~\cite{equitzcsr91} which works at a coarse description distortion
$D_k^\prime$, and fine description distortion of $D_k$.
These give the corresponding terms in the expressions for $D_w$ and $D_s$
respectively in the theorem. The bitrate for the coarse description which
will be made available to both the users is
\[\sum_{k=K^\prime+1}^K \frac{1}{2}\log\frac{\sigma_k^2}{D_k^\prime},\]
and the refinement layer which will be made available only to the strong
user has a bitrate of
\[\sum_{k=K^\prime+1}^K
\frac{1}{2}\log\frac{D_k^\prime}{D_k}.\]

Thus over all we need to send bits at a rate of
\begin{align} \sum_{k=1}^L \frac{1}{2}\log\frac{\sigma_k^2}{D_k^{\prime}} +
\sum_{k=K^\prime+1}^K \frac{1}{2}\log\frac{\sigma_k^2}{D_k^\prime}
\label{hybrid.eq:weakdigitalsourcerate}
\end{align}
to both the users, and in addition bits at the rate of
\begin{align}
 \sum_{k=1}^L
\frac{1}{2}\log\frac{D_k'}{D_k}
+ \sum_{k=K^\prime+1}^K
\frac{1}{2}\log\frac{D_k^\prime}{D_k}
\label{hybrid.eq:strongdigitalsourcerate}
\end{align}
to the strong user. In the next two steps we show how this is accomplished.

\paragraph{Sub-channels $L+1$ through $K^\prime$} As discussed above, on
the $m$-th sub-channel a power of $P_m'-P_m''$ is used for
uncoded transmission. The rest of the power is allocated as follows:
$P_m-P_m'$ is used for sending bits to both the receivers. The rest of
the power $P_m''$ is used to send bits which will be
decoded only the strong receiver. When decoding the common bits, both
receivers treat the rest of the power as interference. Hence the bitrate
of the common part is limited by the weaker user resulting in a rate of
\[ \sum_{m=L+1}^{K^\prime}
\frac{1}{2}\log\left(1+\frac{P_m-P_m'}{P_m'+N_w}\right).\]
As mentioned earlier, upon decoding, both users will subtract the codeword
corresponding to the decoded common bits from their received signal. To
send additional bits to the strong user, we use the concept of dirty-paper
coding~\cite{costawd83}\cite{gelfandpcc80}. The uncoded transmission can be
thought of as Gaussian side-information (or ``dirt'') which is known at the
encoder. From Costa~\cite[Sec. II]{costawd83}, we know that a power
allocation of $P_m''$ can support a rate of
\[ \sum_{m=L+1}^{K^\prime}
\frac{1}{2}\log\left(1+\frac{P_m''}{N_s}\right)\]
to the strong user.

\paragraph{Sub-channels $K^\prime+1$ through $M$} In these sub-channels no
uncoded transmission is performed. We use the idea of superposition
coding~\cite{coverbc72} to deliver a common bitstream to both the users and
in addition a refinement bitstream to only the stronger user. With a power
allocation of $P_m-P_m'$ to the common bitstream and the rest 
$P_m'$ to the refinement bitstream, we get the following
bitrates for the common and refinement bit streams respectively
\begin{align*}
\sum_{m=K^\prime+1}^M&
\frac{1}{2}\log\left(1+\frac{P_m-P_m'}{P_m'+N_w}\right)\\
\sum_{m=K^\prime+1}^M&
\frac{1}{2}\log\left(1+\frac{P_m'}{N_s}\right).
\end{align*}

Thus the total rate available for sending a common bitstream to
both the users is
\[ \sum_{m=L+1}^M \frac{1}{2}\log\left(1+\frac{P_m-P_m'}{P_m'+N_w}\right).\]
Condition \eqref{hybrid.eq:stronguserdigitalcond} ensures that this is
sufficient to handle the rate of the common bit stream in
\eqref{hybrid.eq:weakdigitalsourcerate}. Similarly, the total rate
available for sending an enhancement bitstream to the stronger user is
\begin{align*}
\sum_{m=L+1}^{K^\prime} \frac{1}{2}\log\left(1+\frac{P_m''}{N_s}\right) +
\sum_{m=K^\prime+1}^M \frac{1}{2}\log\left(1+\frac{P_m'}{N_s}\right)
\end{align*}
which is larger than the rate of the enhancement bitstream in
\eqref{hybrid.eq:strongdigitalsourcerate} by condition
\eqref{hybrid.eq:stronguserdigitalcond}. This completes the proof.

\end{appendices}

\small
\bibliographystyle{IEEEtran}

\end{document}